\documentclass[unnumsec,webpdf,contemporary,large]{oup-authoring-template}%

\theoremstyle{thmstyleone}%
\theoremstyle{thmstyletwo}%
\newtheorem{example}{Example}%
\theoremstyle{thmstylethree}%

\usepackage{booktabs} 
\usepackage{listings} 
\usepackage{xcolor}

\lstdefinestyle{json}{
  basicstyle=\ttfamily\scriptsize,
  keywordstyle=\color{blue},
  stringstyle=\color{teal},
  showstringspaces=false,
  breaklines=true,
  frame=none,
  aboveskip=0pt,
  belowskip=0pt
}

\lstdefinelanguage{cypher}{
  morekeywords={
    MATCH, WHERE, RETURN, WITH, CREATE, DELETE, SET, MERGE, TYPE, LABELS, SIZE, IN, CASE, WHEN, THEN, ELSE, END, REDUCE, SPLIT,
    ON, OPTIONAL, DETACH, AS, DISTINCT, UNWIND, UNION, ALL,
    CALL, YIELD, LIMIT, SKIP, ORDER, BY, ASC, DESC, TRUE, FALSE, NULL
  },
  sensitive=true,
  morecomment=[l]{//},
  morestring=[b]{"},
  morestring=[b]{'},
}

\begin{document}

\journaltitle{Journal Title Here}
\DOI{DOI HERE}
\copyrightyear{2022}
\pubyear{2019}
\access{Advance Access Publication Date: Day Month Year}
\appnotes{Paper}

\firstpage{1}


\title[Enhancing RNA-KG with Properties]{RNA-KG v2.0: An RNA-centered Knowledge Graph with Properties}

\author[1,$\ast$]{Emanuele Cavalleri\ORCID{0000-0003-1973-5712}}
\author[1]{Paolo Perlasca\ORCID{0000-0001-6674-2822}}
\author[1,2,$\ast$]{Marco Mesiti\ORCID{0000-0001-5701-0080}}
\authormark{Cavalleri et al.}

\address[1]{\orgdiv{Computer Science Department}, \orgname{University of Milan}, \orgaddress{\street{Via Celoria 18}, \postcode{20122}, \state{Italy}}}
\address[2]{\orgdiv{Environmental Genomics and Systems Biology Division}, \orgname{Lawrence Berkeley National Laboratory}, \orgaddress{\postcode{94720}, \state{CA}, \country{USA}}}
\corresp[$\ast$]{Corresponding authors. \href{email:emanuele.cavalleri@unimi.it}{emanuele.cavalleri@unimi.it} \href{email:marco.mesiti@unimi.it}{marco.mesiti@unimi.it}} 

\received{Date}{0}{Year}
\revised{Date}{0}{Year}
\accepted{Date}{0}{Year}



\abstract{
RNA-KG is a recently developed knowledge graph that integrates the interactions involving coding and non-coding RNA molecules extracted from public data sources. It can be used to support the classification of new molecules, identify new interactions through the use of link prediction methods, and reveal hidden patterns among the represented entities. In this paper, we propose RNA-KG v2.0, a new release of RNA-KG that integrates around $100M$ manually curated interactions sourced from 
$91
$ linked open data repositories and ontologies.
Relationships are characterized by standardized properties that capture the specific context (e.g., cell line, tissue, pathological state) in which they have been identified.
In addition, the nodes are enriched with detailed attributes, such as descriptions, synonyms, and molecular sequences sourced from platforms such as OBO ontologies, NCBI repositories, RNAcentral, and Ensembl. The enhanced repository enables the expression of advanced queries that take into account the context in which the experiments were conducted.
It also supports downstream applications in RNA research, including ``context-aware'' link prediction techniques that combine both topological and semantic information. 
}


\maketitle

\section{Introduction}

RNA molecules play a pivotal role in various physiological and pathological processes, as confirmed by extensive research conducted over the past two decades~\cite{Bartel2004,Cech2014,guttman_rinn_2012}. Beyond their fundamental role in gene expression, these molecules have emerged as promising tools to provide new insights in the treatment of cancer, genetic and neurodegenerative disorders, cardiovascular and infectious diseases~\cite{damase21}. 
Several genomics laboratories release datasets containing interactions between RNAs and other bio-entities such as genes, proteins, chemicals, diseases, and phenotypes. However, the integration and standardization of these data remain a significant challenge, since a common identification scheme is missing to represent RNA molecules and their relationships, and manual efforts are often required to retrieve standard identifiers. 

RNA-KG~\cite{rnakg} is the first ontology-based knowledge graph (KG) that gathers biological knowledge about RNAs from more than 60 public databases, integrating more than $12.5M$ functional relationships with genes, proteins, chemicals, and other ontologically grounded biomedical entities.
RNA-KG is a core project of the Monarch Initiative KG-Hub~\cite{kghub}, a platform that enables standardized construction, exchange, and reuse of KGs.
During the construction of RNA-KG we have noticed that RNA-related relationships frequently depend on: $i)$ the sequence, structure, and/or the genomic loci the RNA is transcribed from; and, $ii)$ on the cell line, tissue, or more in general the ``context'' where they occur. Introducing this information in the KG would be an enhancement for obtaining a more sophisticated representation of the domain and could be used in several downstream analyses and applications. 

This paper introduces RNA-KG v2.0, a new version of RNA-KG that has been extended from multiple viewpoints. 
First, relationships from 20 new data sources have been integrated, and those already stored have been more thoroughly characterized. For example, the ``generic'' {\em RNA-causes or contributes to condition-disease} has now been detailed in specific instances (e.g., {\em RNA-over-expressed in-disease} and {\em RNA-under-expressed in-disease}) according to the metadata provided by the sources. 
Moreover, a new identification scheme for RNA molecules, based on RNAcentral~\cite{rnacentral} for non-coding RNAs and Ensembl~\cite{ensembl} for coding transcripts, is now adopted. 
This scheme enables the representation of transcript variants (i.e., isoforms), overcoming the limitations of previous releases, which collapsed multiple isoforms into a single gene-level entity. For example, the long non-coding RNAs LINC-PINT-205 and LINC-PINT-206 previously collectively identified through the Entrez identifier {\it Entrez:378805?lncRNA} are now represented as  {\it URS00025E6A5F\_9606} and {\it URS00005A4FBF\_9606}.
RNA entities in RNA-KG are also enhanced by integrating the Rfam~\cite{rfam} node classification scheme, which improves the accuracy and granularity of RNA categories.
Finally, properties for both entities and relationships are now included in the knowledge base. This information is of paramount importance for discriminating bio-entities and relationships based on their data sources and the ``context'' where they have been acquired.
For instance, the RNA regulation of a gene may depend on specific sequence motifs. RNA isoforms may differ by a few nucleotides, but one may bind to a protein-coding-gene due to the presence of the motif, while the other does not. Similarly, the over-expression of an RNA might be associated with Alzheimer's disease in brain tissues but not in the blood. This information can now be exploited for conducting different kinds of analysis and developing meaningful applications. 

The resulting KG integrates entities and relationships from 80 
publicly available repositories and 11 biomedical ontologies with a total of
$6,553,767$ nodes involved in $99,936,712$ relationships. Nodes and edges are characterized by $27,242,075$ and $242,896,588$ properties.
The new RNA-KG version is available as a property graph in Neo4j~\cite{neo4j}. 
A web portal and an API have been developed to support the users in accessing and interacting with the information stored in the knowledge base. By means of the web portal, interested users can decide to generate {\it views} (i.e. subgraphs of interest~\cite{torgano}) by exploiting the end-point or download {\it views} already extracted by our team. Through the API and thanks to the adoption of standard identification schemes, the integration with the information provided by other portals (like RNAcentral) is made easy and further kinds of analysis can be carried out.

The paper is organized as follows. The first section introduces the new considered data sources and the main modifications to the proposed identification scheme.
The second and third sections are devoted: $i)$ to presenting the characteristics of the included bio-entities and relationships along with their properties; $ii)$ to providing an example of the use of the knowledge graph. The fourth section describes the characteristics of the web portal and the API. 
Finally, use cases and applications are discussed where RNA-KG properties can be exploited to show the relevance of our knowledge base.

\section{New Data Sources and Identification Schemes}\label{sec:rnakgnewrelease}

RNA-KG v2.0 integrates publicly available relationships involving biomedical entities from  $91$ data sources and ontologies.
Twenty new sources were identified: RNAcentral, Ensembl, DisGeNET~\cite{disgenet}, GeneMANIA~\cite{genemania}, CTD~\cite{ctd}, ClinVar~\cite{clinvar}, STRING~\cite{string}, starBase2~\cite{starbase2}, microT~\cite{microt}, The Human Protein Atlas~\cite{hpa}, miRanda~\cite{miranda}, Reactome~\cite{reactome}, HGNC~\cite{hgnc}, GTEx~\cite{gtex}, POSTAR2~\cite{postar2}, UniProtKB~\cite{uniprot}, RNAhybrid~\cite{rnahybrid}, PhenomiR~\cite{phenomir}, COSMIC~\cite{cosmic}, and circBase~\cite{circbase}
. 
These sources were considered because they $(i)$ increase the coverage of interactions that describe the ``RNA world'' (e.g. {\em RNA-ribosomally translates to-protein} interactions from Ensembl); $(ii)$ are well-reputed sources that do not or marginally deal with RNA molecules but enhance the KG quality, including relationships involving entities that are connected to RNA molecules in RNA-KG (e.g. {\em protein-molecularly interacts with-protein} interactions from STRING). Supplementary Tables S1-S3 
detail the new relationships included along with the entities involved.

Eleven standard biomedical ontologies are used to establish common semantics: Human Phenotype Ontology (HPO~\cite{hpo}), Gene Ontology (GO~\cite{go})
Monarch Merged Disease Ontology (Mondo~\cite{mondo}), Vaccine Ontology (VO~\cite{vo}), Chemical Entities of Biological Interest (ChEBI~\cite{chebi}), Uber-anatomy Ontology (Uberon~\cite{uberon}), Cell Line Ontology (CLO~\cite{clo}), PRotein Ontology (PRO~\cite{pro}), Sequence Ontology (SO~\cite{so}), Pathway Ontology (PW~\cite{pw}), and Relation Ontology (RO~\cite{ro}).
We selected these ontologies because their terms, properties, and hierarchical structures are commonly accepted by the scientific community to unequivocally describe biological classes and entities such as diseases, phenotypes, chemicals, biological processes, proteins, and relations between them.
Bio-entities that are not currently represented in one of these ontologies are modeled using well-accepted terminologies. They include DrugBank accession numbers for identifying drugs, NCBI Entrez identifiers for genes and pseudogenes, RNAcentral identifiers for ncRNA molecules, Ensembl transcript identifiers for mRNA sequences, NCBI dbSNP~\cite{dbsnp} rsIDs and COSMIC identifiers for germline and somatic variants.
Grounding RNA molecules in RNAcentral and Ensembl identifiers allows the precise RNA sequence to be distinguished among RNA isoforms. 
Table~\ref{tab:referenceschema} details the most relevant identification schemes for the main bio-entities according to the proposed histograms. 
Look-up tables were used to map proprietary identifiers to standard identification schemes as shown 
in~\cite{rnakg}.

\begin{table}[t]
    \centering
    \begin{tabular}{|l|l|l|}
    \hline
         {\bf Entity} & {\bf Scheme} & {\bf Example} \\\hline
         Variant & \begin{tabular}[x]{@{}l@{}}dbSNP\\COSMIC\end{tabular} & \begin{tabular}[x]{@{}l@{}}rs766102409\\COSV60127483\end{tabular} \\\hline
         ncRNA & RNAcentral & URS00000478B7\_9606\\\hline
         mRNA & Ensembl & ENST00000713680 \\\hline
         circRNA & circBase & hsa\_circ\_0018046 \\\hline
         Chemical & ChEBI & CHEBI:4021 \\\hline
         Protein & PRO 
         & PR:Q8WXF3 \\\hline
         Gene & Entrez & Entrez:1954 \\\hline
         Cell & CLO & CLO:0003725 \\\hline
         GO term & GO & GO:0140657 \\\hline
         Disease & Mondo & MONDO:0020683 \\\hline
         Phenotype & HPO & HP:0040064 \\\hline
         Anatomy & Uberon & UBERON:0002169 \\\hline
         Vaccine & VO & VO:0010137 \\\hline
         Pathway & \begin{tabular}[x]{@{}l@{}}PW\\Reactome\\WikiPathways
         \end{tabular} & \begin{tabular}[x]{@{}l@{}}PW:0000035\\R-HSA-9837999\\WP5090\end{tabular} \\\hline
    \end{tabular}
    \caption{Identification schemes for the main bio-entities.}
    \label{tab:referenceschema}
\end{table}

\begin{figure}[t]
\includegraphics[width=0.45\textwidth]{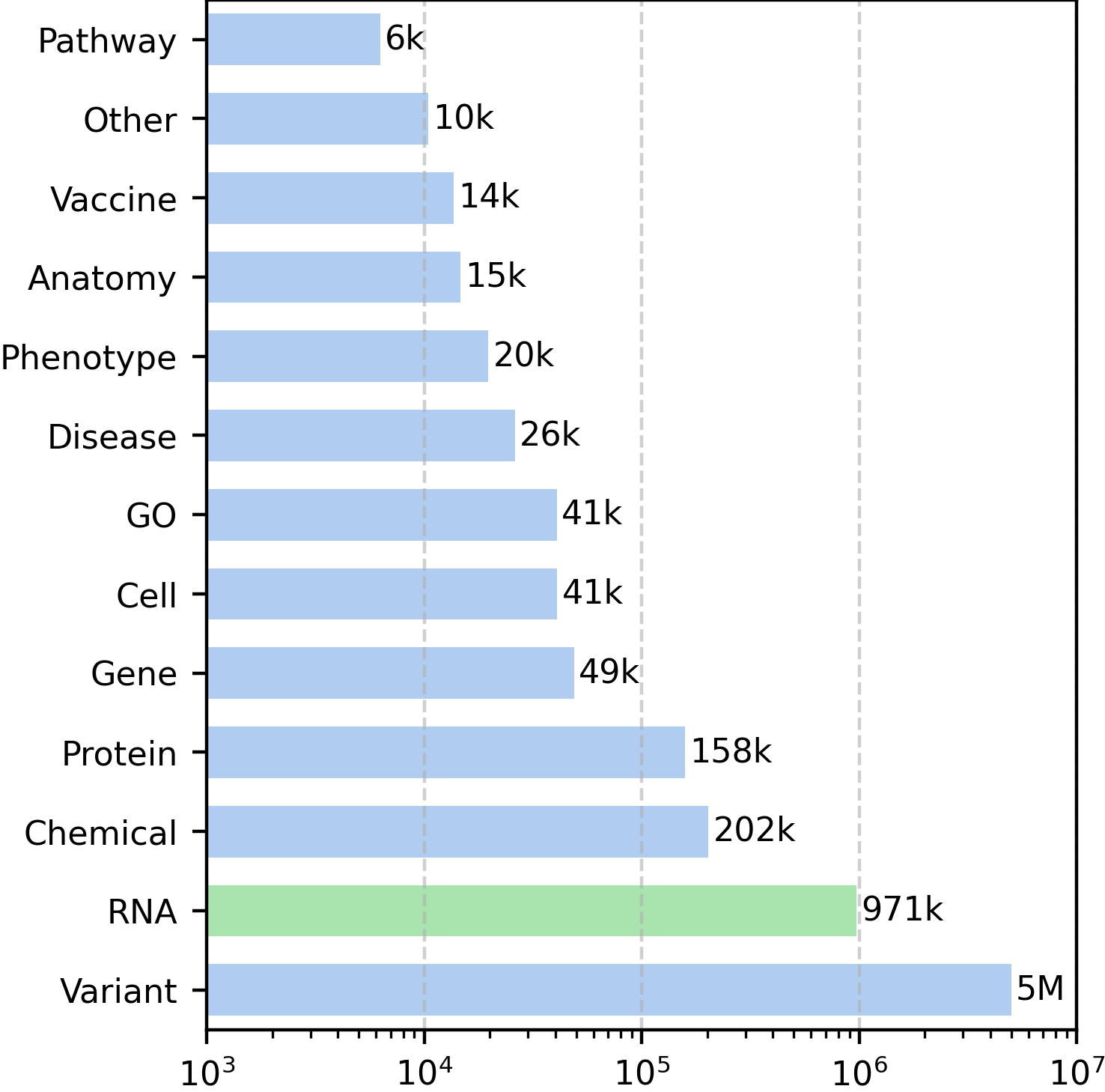}
\caption{Node type distribution.}
    \label{fig:pie_ont}
\end{figure}

\section{RNA-KG Bio-Entities} 
\label{sec:nodes}

\subsection{Node type distribution}

RNA-KG includes approximately $6.5M$ typed nodes with $139$ distinct node types. Ontology classes are used to define node types (e.g. {\it SO:0000276} for {\tt miRNA}).
Additional node types have also been introduced by parsing ontologies to enhance entity types with specific relevance within the ``RNA world''. For instance, the type {\tt Cardiovascular disease} was assigned to specific disease entities based on hierarchical relationships within Mondo and the {\tt Transcription Factor} ({\tt TF}) type has been associated with proteins whose label in PRO contains the string ``transcription factor'' or matches known TF symbols from curated sources.
This fine-grained representation facilitates the construction of domain-specific {\it views} using Cypher~\cite{cypher} queries or the graphical interfaces of the web portal.

\begin{figure}[b]
\includegraphics[width=0.45\textwidth]{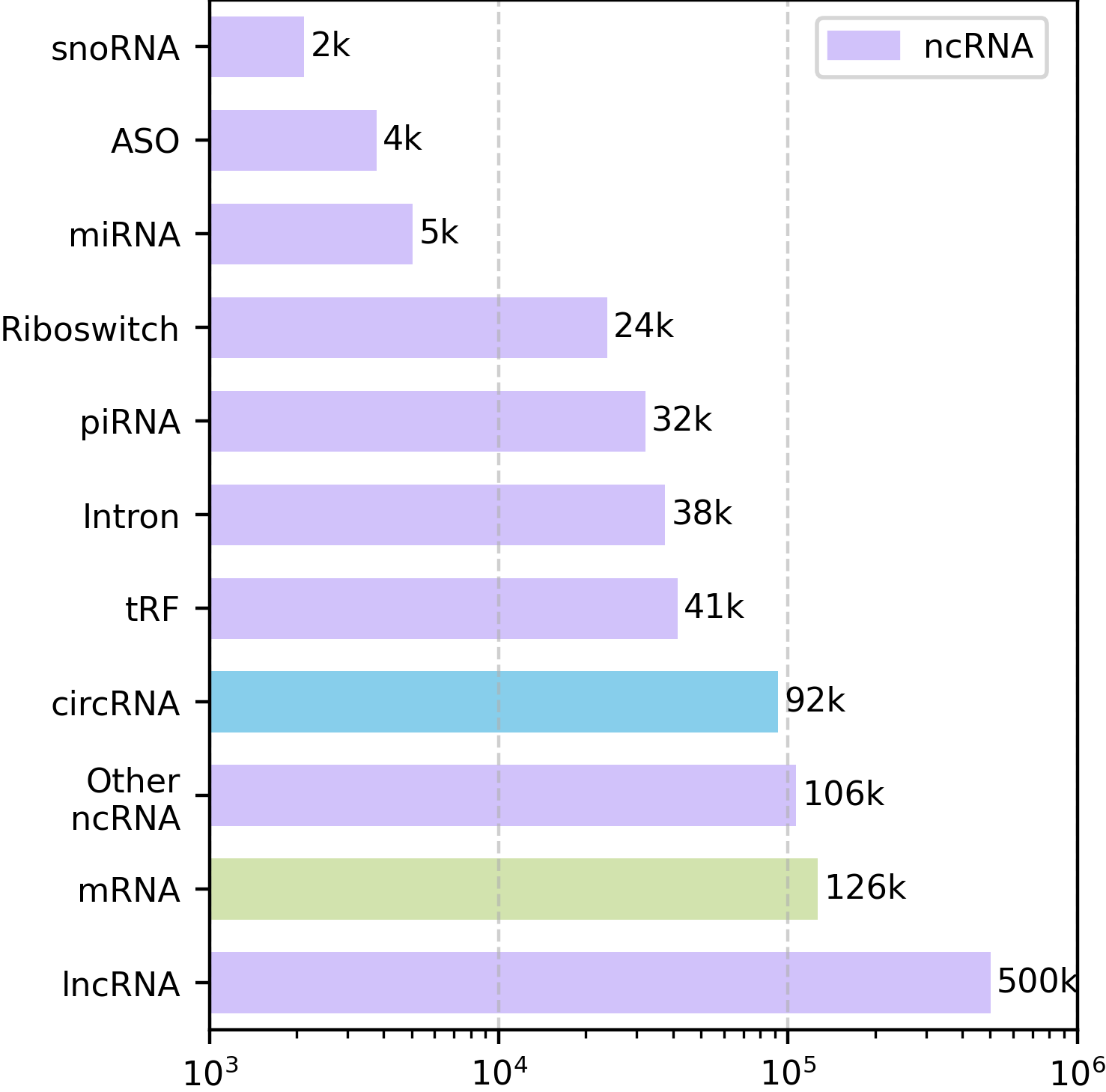}
    \caption{RNA node distribution.}
    \label{fig:rna_nodes}
\end{figure}

Fig.~\ref{fig:pie_ont} shows the node types distribution in RNA-KG.
The most represented class is {\tt Variant} ($\approx5M$ nodes) due to the integration of large-scale variant databases (e.g. dbSNP and COSMIC) that catalog millions of mutations across the human genome. Note that each known mutation event (e.g. an A-to-T substitution occurring in a specific genomic locus) leads to the creation of a new variant identifier. The second most common node type is {\tt RNA} ($\approx971k$ nodes). RNA sequences are typed using SO classes and Rfam categories such as {\tt mRNA}, {\tt lncRNA}, and {\tt miRNA}. Fig.~\ref{fig:rna_nodes} details the distribution of RNA types. Non-coding RNAs represent a substantial portion of this set ($\approx752k$ nodes), as supported by current estimates~\cite{rfam,rnacentral}.
mRNAs account for around $126k$ nodes, a number consistent with the expected diversity of transcript isoforms derived from the $\approx20$–$22k$ protein-coding genes in the human genome~\cite{Salzberg2018}.

\subsection{Bio-entities' properties}

Properties for RNA-KG nodes correspond to the attributes associated with the ontological terms.  
Among the available attributes, {\tt label}, {\tt description}, and {\tt synomyms} are associated with nodes that represent amino acids, anatomical parts, biological processes, biological roles, cells and cell lines, cellular components, chemicals, diseases, genomic features (e.g. histone modifications and sequence locations), lipids, molecular functions, pathways, phenotypes, proteins, species, and vaccines.
Additional attributes are included for specific nodes' categories. Specifically, {\tt charges}, {\tt masses}, {\tt formulae}, {\tt SMILES structures}, {\tt CAS numbers}, and {\tt InChIKeys} for chemical entities; amino acid {\tt sequences} and {\tt species} for proteins.

\begin{figure}[t]
\includegraphics[width=0.45\textwidth]{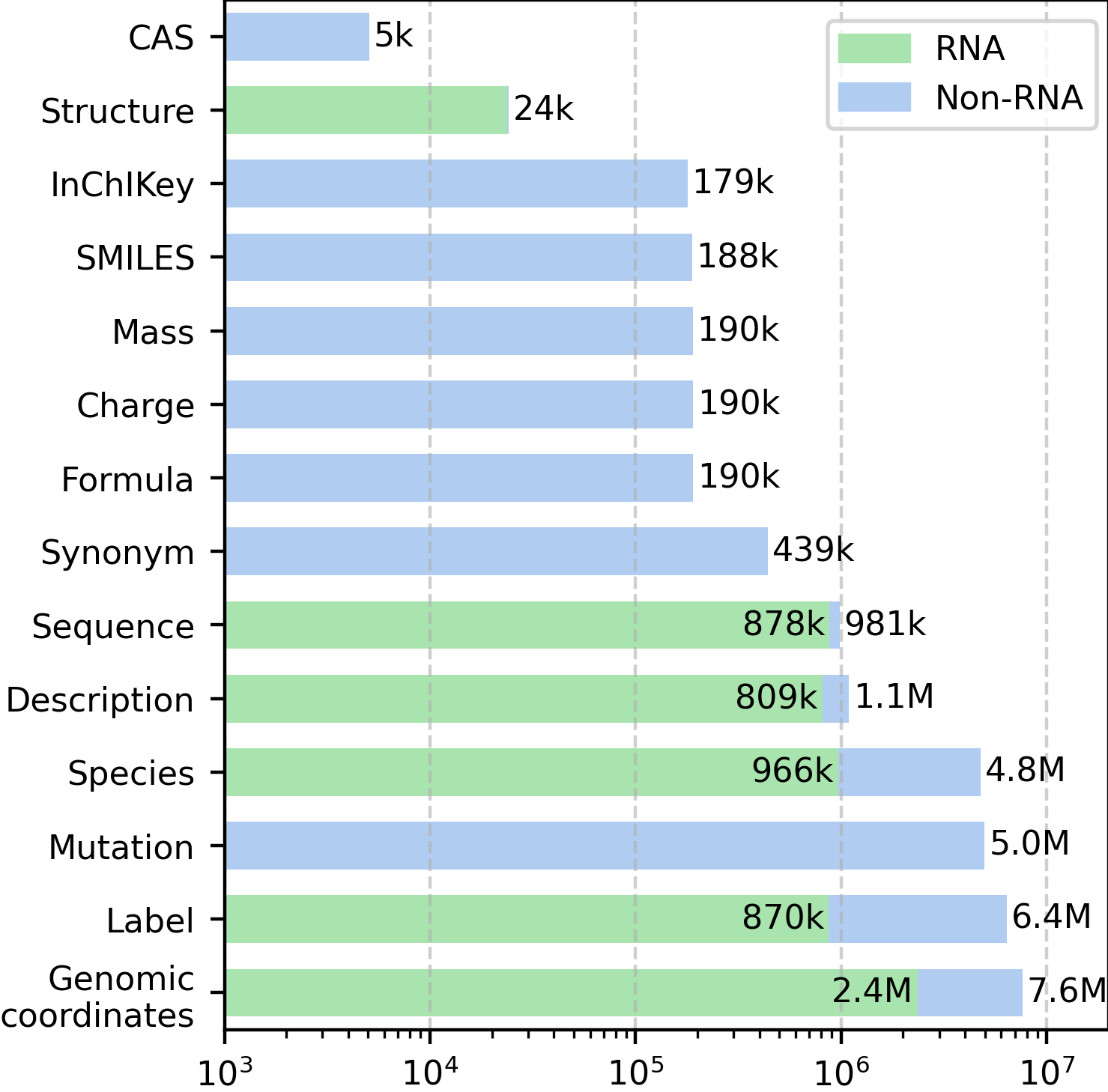}
    \caption{Node property distribution.}
    \label{fig:rna_properties}
\end{figure}

On the other hand, RNA-KG nodes representing entities that cannot be directly mapped to ontological terms (e.g. genes, RNAs, and variants) are characterized by a subset of the attributes that can be retrieved from the corresponding terminologies.
Specifically, $(i)$ DrugBank provides {\tt labels}, {\tt sequences}, {\tt synonyms}, {\tt InChIKeys}, and {\tt CAS numbers} for drugs;
$(ii)$ NCBI Entrez provide {\tt labels}, {\tt species}, {\tt synonyms}, {\tt descriptions}, and {\tt genomic coordinates} with associated nucleotide {\tt sequences} for genes; $(iii)$ NCBI dbSNP and COSMIC provide {\tt labels}, {\tt species}, and {\tt genomic coordinates} with associated {\tt mutations} 
that occur in the nucleotide chain.
Many efforts have been devoted to produce uniform representation of specific attributes like {\tt structures}, {\tt sequences}, {\tt mutations}, and {\tt genomic coordinates}.
Fig.~\ref{fig:rna_properties} shows the distribution of nodes' properties in RNA-KG, highlighting the cardinality of RNA nodes' properties, which will be detailed in the following.

\subsection{Properties specific for RNA molecules}\label{sec:rnaproperties}

RNA molecules are characterized by a set of properties that include {\tt labels}, {\tt structures}, {\tt descriptions}, {\tt genomic coordinates}, and 
{\tt sequences}. These properties are retrieved primarily from centralized repositories such as Ensembl and RNAcentral, minimizing the risk of heterogeneity in annotation. The property {\tt genomic coordinates} is the most numerous since it is a list of genomic loci (whose items are provided with the standard format {\it chromosome: start-end strand}) from which RNA molecules are transcribed (the same RNA sequence can present multiple loci in the genome).
Proprietary specialized databases contribute additional annotations when molecules cannot be mapped on Ensembl/RNAcentral terminologies (details in the Supplementary material).

\begin{figure*}
    \centering
    \includegraphics[width=0.8\linewidth]{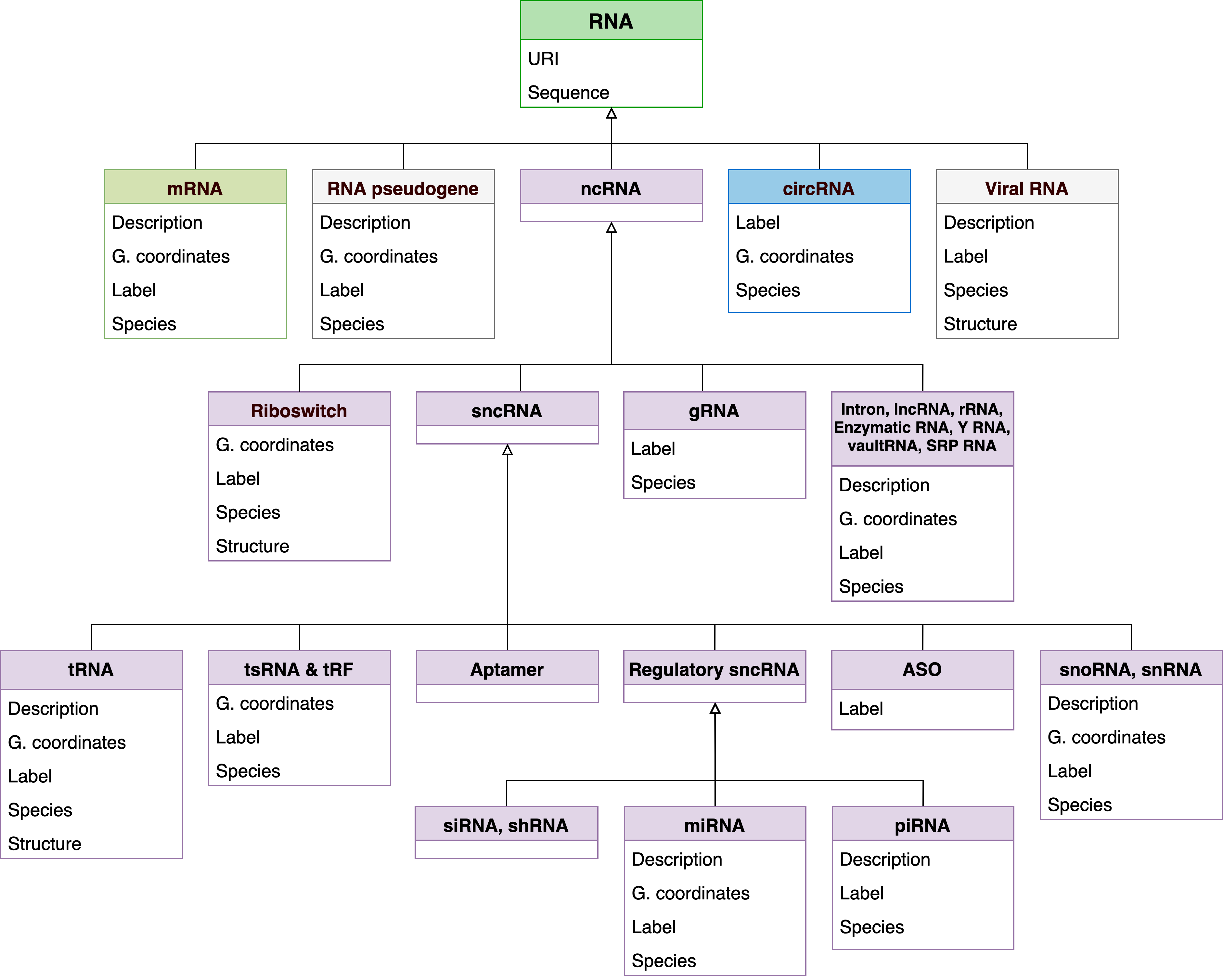}
    \caption{Hierarchy of RNA molecules with associated properties.
    }
    \label{fig:rnaproperties}
\end{figure*}

Fig.~\ref{fig:rnaproperties} shows the hierarchical organization of RNA molecules along with the associated properties. A property is added to an RNA class if at least one instance of that class is enriched with that property. Child nodes inherit parent nodes' properties.
For the sake of readability, RNA drugs and many RNA (sub)categories that are under-represented in RNA-KG are removed from the visualization. For example, RNA antisense oligonucleotides subclasses (e.g. PMO ASO, 2MeO ASO), tRF subtypes (e.g. tRF-1, tRF-3), scaRNAs, T-box riboswitches, ribozymes and RNA components of RNAses, retained introns, precursor RNAs, bacterial RNAs, and RNA decay molecules are removed because they do not add relevant properties to the hierarchy. 
It is worth noting that nucleotide sequences are always associated with an RNA class but some of them miss the associated genomic coordinates because they are synthetic (e.g. siRNAs, shRNAs, ASOs) or they are co-located with multiple transcripts, as for piRNAs, and no certain coordinates are provided by the original source. 
Viral RNAs (that usually represent the entire genome of the pathogen since RNA virus genomes usually contain relatively few genes~\cite{viralrna}), bacterial riboswitches, and tRNA sequences are enriched with their secondary {\tt structures} (expressed using the base pair ``dot-bracket notation''~\cite{dotbracket}). These foldings are crucial for RNA functional analysis as they influence RNA stability, interactions with proteins or other RNAs, and the regulation of gene expression~\cite{rnafolding}.

\section{RNA-KG Edges}

\subsection{Edge type distribution}

RNA-KG includes approximately $100M$ typed edges with $424$ distinct edge types. 
Edge types are grounded in OBO ontologies properties. Specifically, we used RO properties for annotating relationships identified within linked open data sources.
The hierarchical organization of concepts in RO allows the expression of different kinds of edges at different granularities. For example, the general property {\it RO:0002434} {\em interacts with}
(the most numerous edge type in RNA-KG due to the integration of homogeneous sources such as RNAInter~\cite{rnainter}) can be substituted with more specific properties such as {\em molecularly interacts with} or {\em genetically interacts with}.
Inverse relationships are included in RNA-KG (e.g. {\em location of} and its inverse {\em located in}).
Besides edges coming from linked open data, we introduced edges stemming from the $(i)$ integration of the eleven bio-ontologies according to the abstraction strategy described in~\cite{owlnets} and $(ii)$ entity linking technique according to the class-based modeling described in~\cite{callahan23}. The model links entities that are not directly mappable to an ontology term (i.e. database entities without a corresponding ontological identifier) to ontology classes using the {\em rdfs:subClassOf} property.

\begin{figure}[t]
\hspace*{-.7cm}\includegraphics[width=0.5\textwidth]{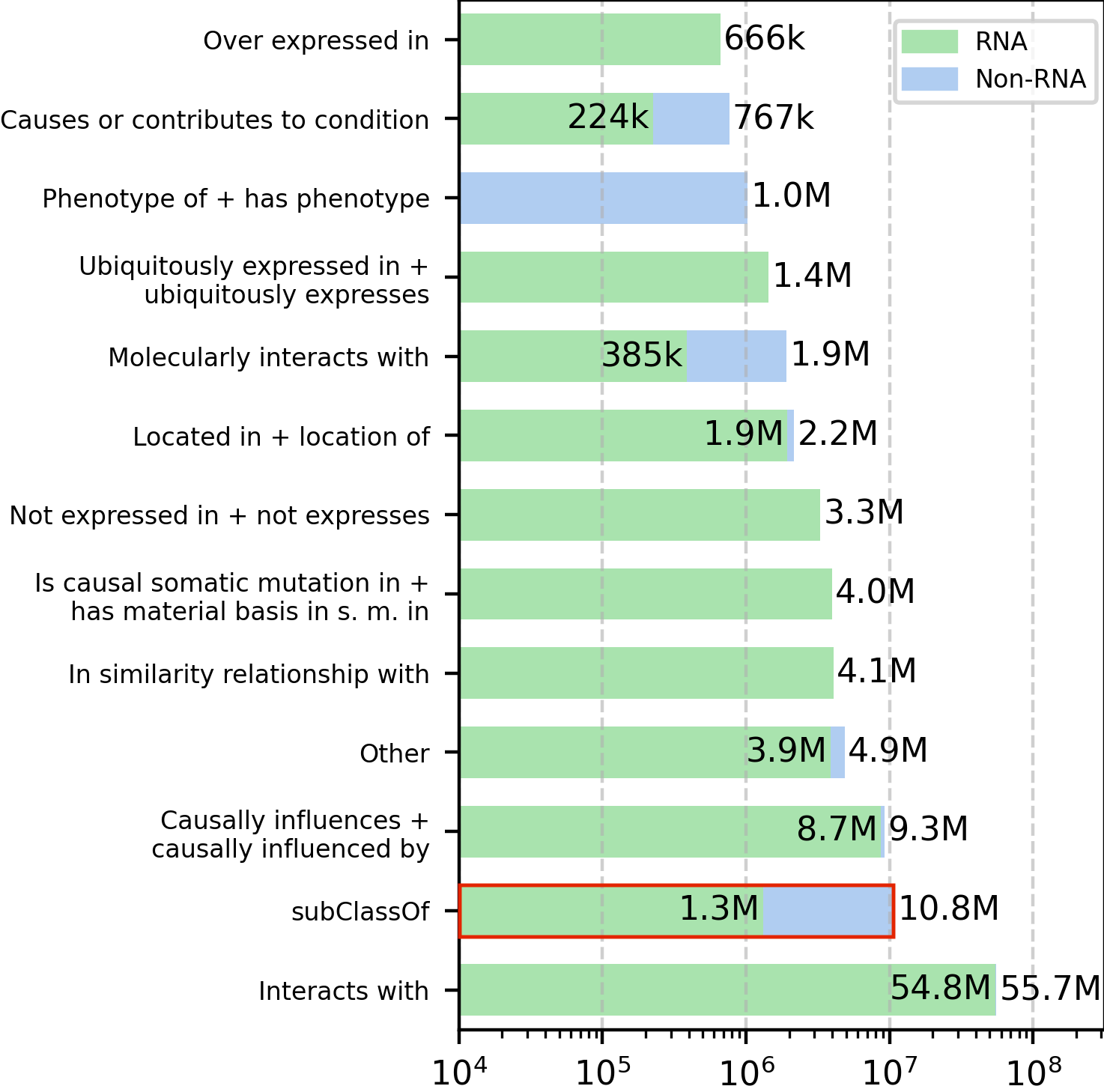}
    \caption{Edge type distribution.}
    \label{fig:edges_barh}
\end{figure}

Fig.~\ref{fig:edges_barh} shows the distribution of edge types in RNA-KG, highlighting the cardinality of the relations involving RNA molecules. The distribution mirrors the biological focus of RNA-KG since more than half of its relationships involve RNAs.
The figure also points out the relevance of {\em rdfs:subClassOf} which can be used for entity linking and ontology integration.

\subsection{Edges' properties}

\begin{figure}[t]
    \centering
    \includegraphics[width=0.45\textwidth]{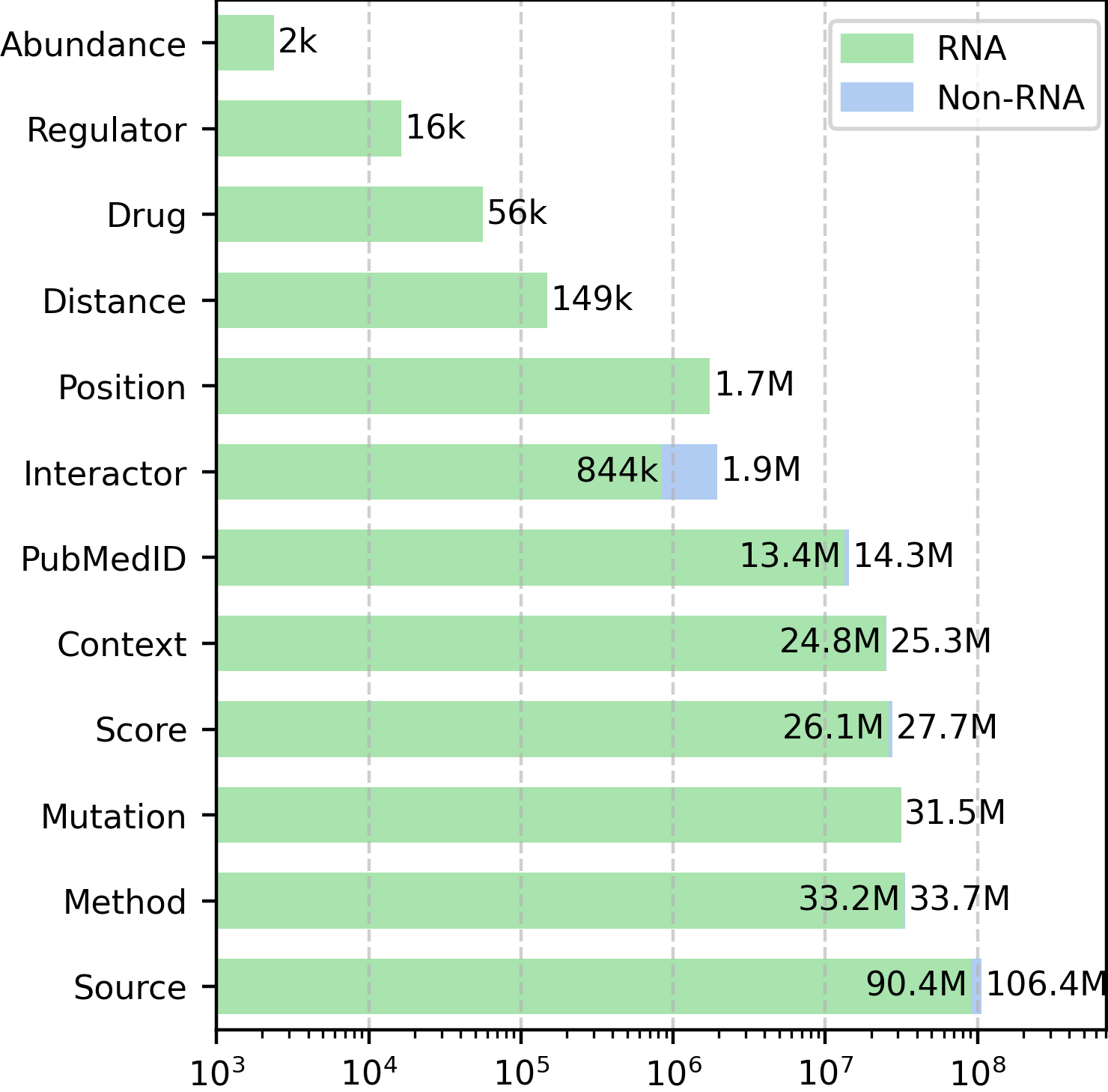}
    \caption{Edge property distribution.}
    \label{fig:edges_properties}
\end{figure}

Fig.~\ref{fig:edges_properties} outlines the most important edges' properties that we have identified in terms of cardinality and biological relevance, highlighting properties involving RNA molecules.
Examples of properties are $i)$ the {\tt score} that indicates the level of confidence for the interaction to occur; $ii)$ the {\tt method} that provides experimental evidence to the interaction (e.g. RT-PCR, AGO-CLIP, ImmunoBlot); $iii)$ the specific tissue, cell line, disease, phenotype (in general, the {\tt context}) where the interaction was analyzed; $iv)$ the {\tt PubMed identifier} of the article where the interaction is described.
Interactions can also be enhanced by {\tt variants}, {\tt epigenetic modifications}, competing molecule or binding partner (RNAs, genes etc., in general, the {\tt interactor}) that provides further context for the documented relationship as described by the original source.
Each relationship within RNA-KG is also associated with the list of {\tt sources} providing the interaction.
Moreover, edges' properties are grounded in bio-ontologies and terminologies, as we have done for nodes and edge types.
For example, {\tt methods} are grounded in the National Cancer Institute Thesaurus (NCIt~\cite{ncit}), {\tt mutations} in dbSNP and COSMIC, and {\tt regulators} in PRO. 

Since we are aware that binary relationships usually adopted in the representation of KG are not adequate to model dependencies between properties (e.g. the dependence of mutations on the context) and ternary relationships (e.g. those observed in ceRNA interactions~\cite{cerna}), we have chosen to represent this information as a list of attributes because it offers a practical compromise between expressivity and simplicity. Modeling these complex dependencies explicitly would require blank nodes or reification patterns, which would increase both the complexity of the graph schema and the computational overhead in querying and maintenance. Instead, by means of the proposed representation, we can easily extract through Cypher queries the subgraphs containing the information required for conducting a certain kind of analysis and thus simplify the application of ML approaches, especially those that are graph-based.

\begin{figure*}[ht]
    \centering
    \includegraphics[width=.9\textwidth]{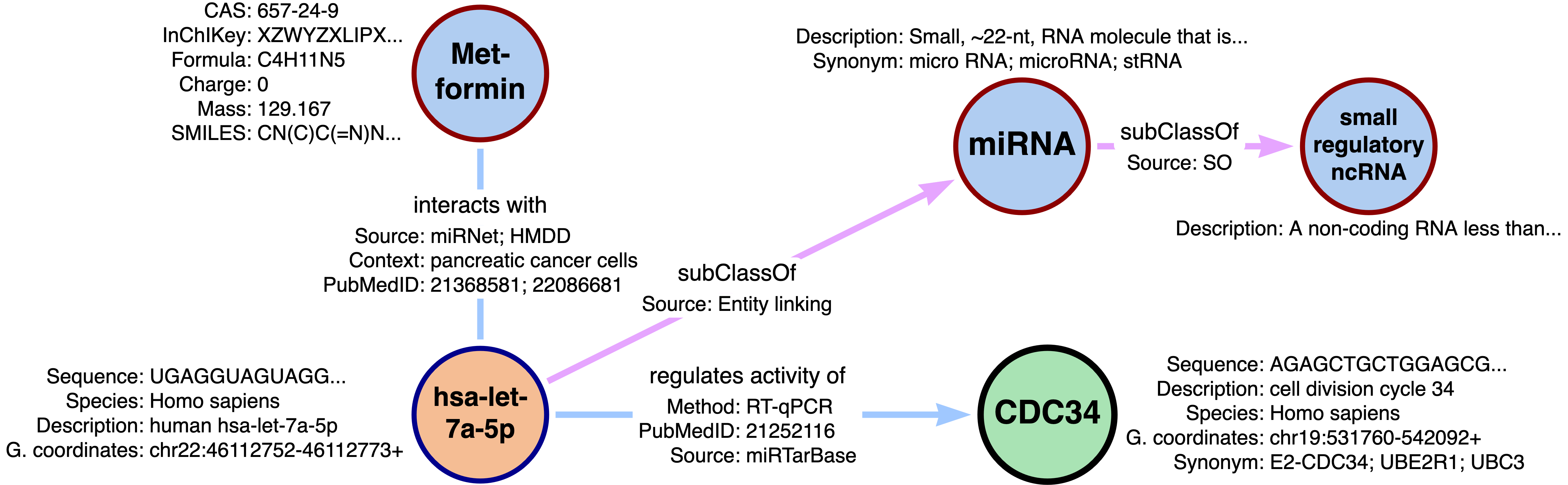}
    \caption{Excerpt of a RNA-KG subgraph.}
        \label{fig:example}
\end{figure*}


\begin{example} 
To illustrate the expressive power of the property-graph edition of RNA-KG, consider the subgraph  in Fig.~\ref{fig:example} depicting the interactions among the miRNA {\it hsa-let-7a-5p}, the protein-coding gene {\it CDC34}, and the small compound {\it metformin} (for the sake of readability, labels are reported instead of identifiers).
Node properties provide biological information to be exploited in downstream analyses. {\tt Species} and {\tt description} are associated with miRNAs and genes and facilitate the disambiguation, curation, and filtering of nodes by organism or functional role. {\tt Genomic coordinates} and {\tt sequence} can support sequence-based analyses such as RNA-target binding prediction, isoform comparison, and integration with data coming from -omic studies.
{\it Metformin} is annotated with attributes derived from the ChEBI ontology, including {\tt formula}, {\tt SMILES}, {\tt mass}, and {\tt charge}. Combining these features with topological metrics (e.g. node degree, centrality, closeness) enables chemo-informatics analyses that integrate chemical and network-based evidence to prioritize drug-like molecules.

Edge properties enrich the semantics of interactions. The edge {\em regulates activity of} between {\it hsa-let-7a-5p} and {\it CDC34} includes the properties {\tt method}, {\tt PubMedID}, and {\tt source}. These attributes assess the reliability of interactions and can be used in interaction-ranking applications~\cite{rank}.
The edge {\it hsa-let-7a-5p} {\em interacts with} {\it metformin} presents the {\tt context} (in the running example, the cancer cells where the interaction was observed) and the {\tt sources} confirming the interaction. 
During {\it views} generation, this knowledge can be used to weigh interactions and prioritize them.

RNA-KG v2.0 also integrates entity linking and ontology nodes and edges (pink relations). {\it hsa-let-7a-5p} is linked to its corresponding SO class {\it miRNA} via a {\em rdfs:subClassOf} relation. This allows hierarchical querying in Cypher and the retrieval of entities based on their ontology categories.
The SO hierarchy includes links between ontology terms such as {\it miRNA} and {\it small regulatory ncRNA} that can be leveraged for reasoning tasks. Furthermore, ontology nodes' properties (e.g. {\tt label}, {\tt description}, and {\tt synonyms}) facilitate entity normalization, mapping, and semantic search.
\end{example} 


\section{The Web portal and API}
The new release of RNA-KG is stored as a property graph database.
A public Neo4j endpoint has been realized to query our KG and made available at \url{https://neo4j.biodata.di.unimi.it} with username and password {\it rnakgv20}.
Data and code for reproducing experiments are available in Zenodo and GitHub at the links: \url{https://doi.org/10.5281/zenodo.10078876}; \url{https://github.com/AnacletoLAB/RNA-KG}.

\begin{figure*}[t]
    \centering
    \includegraphics[width=0.9\linewidth]{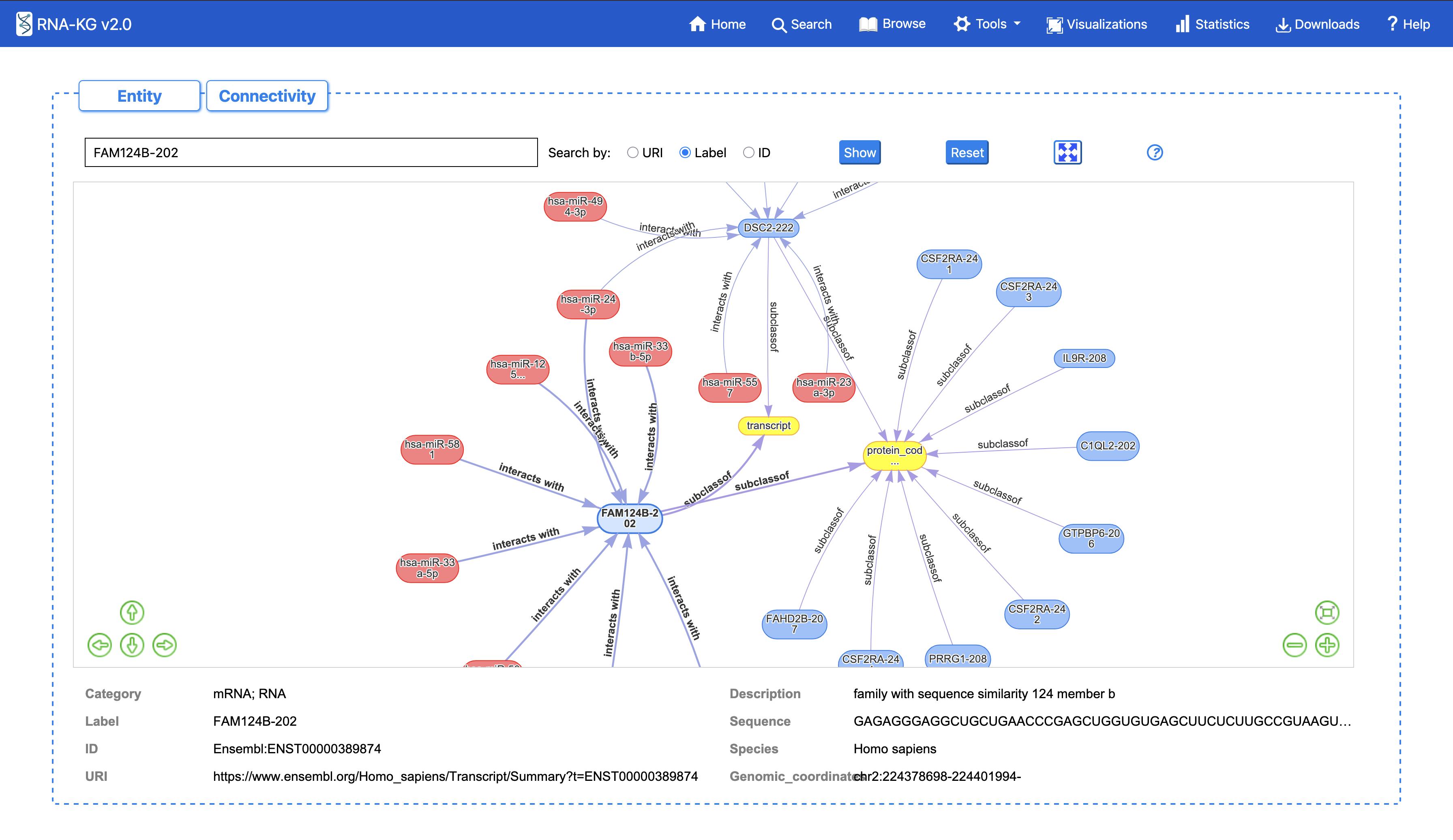}

(a) Exploration facilities

    \centering
    \includegraphics[width=0.9\linewidth]{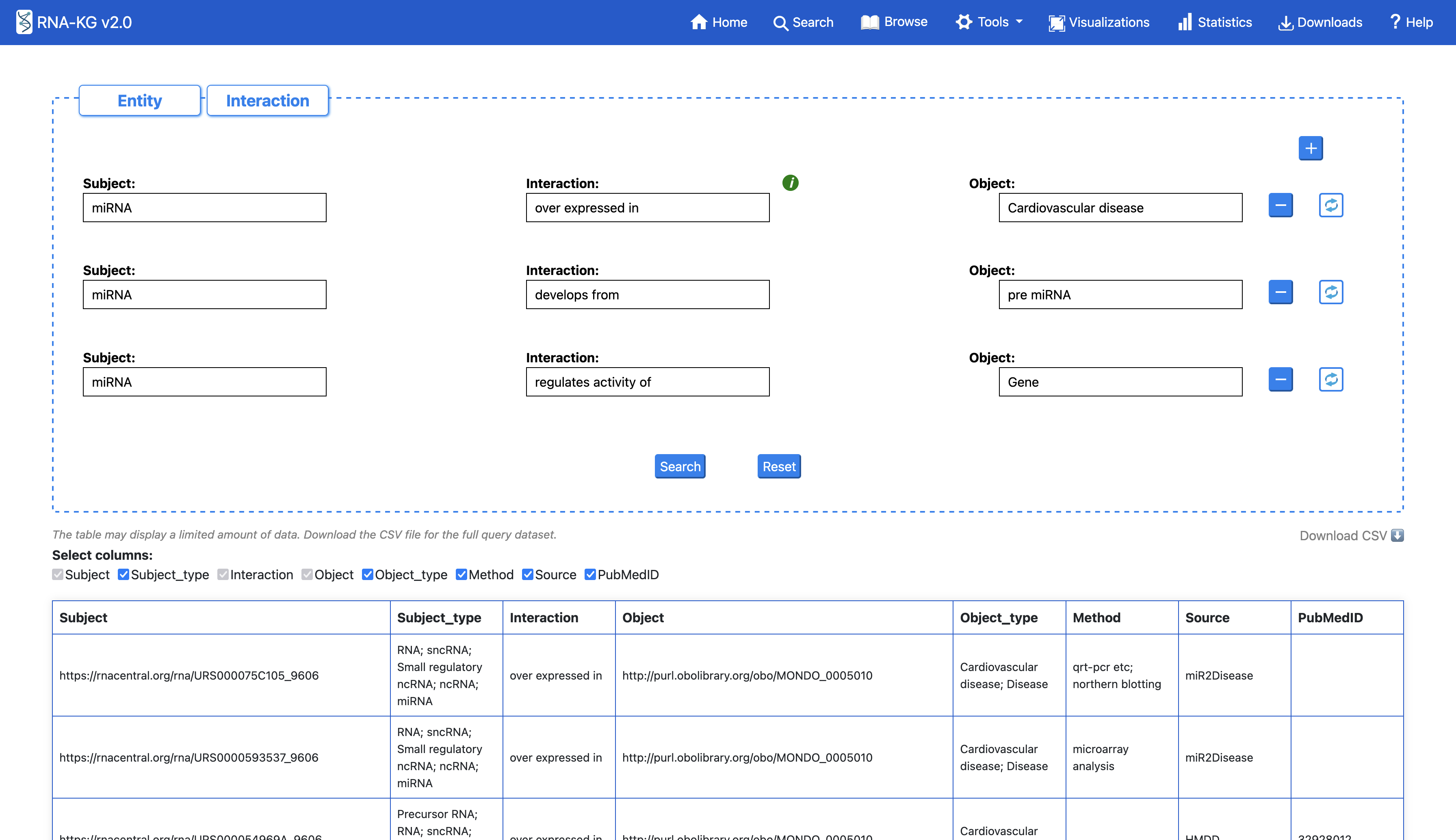}

    (b) Generation of custom {\it views} 
    
    \caption{RNA-KG web portal. 
    }
    \label{fig:website}
\end{figure*}

A web portal is available at \url{https://RNA-KG.biodata.di.unimi.it} and provides several facilities for working with RNA-KG~v2.0. As shown in Fig.~\ref{fig:website}(a), the portal provides a browsing interface for exploring the KG structure. Starting from a node of the graph, it is possible to expand the visualization with its neighboring nodes and see the properties associated with nodes and edges. Moreover, through the interface in Fig.~\ref{fig:website}(b), the user can 
generate a customized {\it view} by specifying the kind of nodes and edges to be included. {\it Views} can be downloaded in CSV. Predefined {\it views} prepared by our team can also be downloaded from the web portal. 

RNA-KG v2.0 is also accessible via a RESTful API that makes it easier the integration with other data sources. The API exposes endpoints to retrieve the properties associated with a given entity, incoming and outgoing relationships for a given entity, to run Cypher queries on the Neo4j database, and to access metadata and schema-level information. 
A list of the main endpoints, including their descriptions, input/output examples, and HTTP methods are provided in Supplementary Table S4
.

\section{Applications and Use Cases}\label{sec:discussion}

RNA-KG v2.0 enhances its utility in multiple ways.
First, it enables fine-grained queries, allowing users to filter interactions not only by molecular entities but also by context, experimental method, or supporting literature.
Moreover, node and edge properties can be used as additional features to improve the performance of the ML models in different tasks. 
All these applications and use cases will be discussed in the remainder of the section.


\subsection{Context-aware Cypher queries}
Filtering on nodes and edges' properties is made easier by expressing fine-grained Cypher queries on RNA-KG. 


\begin{example}\label{ex:1}
    Suppose we wish to retrieve interactions between miRNAs and genes that western blotting assays have experimentally validated. 
    Listing~\ref{lst:query1} shows the Cypher query for retrieving the desired interactions.
\end{example}

\vspace*{-8pt}

\begin{lstlisting}[language=cypher,caption={Query to retrieve interactions between miRNAs and genes validated by a western blotting assay in RNA-KG.}, label={lst:query1}]
MATCH (m:miRNA)-[r]->(g:Gene)
WHERE "western blotting" IN r.Method
RETURN m.URI AS miRNA, TYPE(r) AS interaction_type, r.Method,
    r.PubMedID, r.Source, g.URI AS Gene
\end{lstlisting}

\begin{example}\label{ex:2}
Suppose we wish to retrieve miRNAs that are rich in uracil, which may be relevant for structure or stability studies. 
Listing~\ref{lst:query2} shows a query to retrieve miRNAs whose uracil (``U'') content exceeds 25\%.
The {\it miRNA\_type} property is returned so that users can distinguish mature miRNAs from hairpin precursors (in RNA-KG, hairpin miRNAs are a subclass of miRNAs).
\end{example}

\vspace*{-9pt}

\begin{lstlisting}[language=cypher,caption={Query to retrieve miRNAs whose uracile percentage is above 25\%.}, label={lst:query2}]
MATCH (m:miRNA)
WITH m, REDUCE(u_count = 0, c IN SPLIT(m.Sequence, "") | 
     u_count + CASE WHEN c = "U" THEN 1 ELSE 0 END) AS u_total,
     SIZE(m.Sequence) AS total
WHERE 100.0 * u_total / total > 25
RETURN m.URI AS miRNA, LABELS(m) AS miRNA_type, m.Sequence
ORDER BY miRNA_type
\end{lstlisting}

\vspace*{-9pt}

\begin{example}\label{ex:3}
    Suppose we are interested in retrieving triples of miRNAs, genes, and diseases where both the miRNA-gene and gene-disease interactions share at least one supporting PubMed article. 
    Listing~\ref{lst:query3} shows the corresponding query.
\end{example}

\vspace*{-8pt}

\begin{lstlisting}[language=cypher,caption={Query to retrieve miRNA-gene-disease triples with shared PubMedIDs.
%MM on both interactions.
}, label={lst:query3}]
MATCH (m:miRNA)-[rmg]->(g:Gene)-[rgd]->(d:Disease)
WITH m, g, d,
     [pmid IN rmg.PubMedID WHERE pmid IN rgd.PubMedID] AS common_pmids
WHERE SIZE(common_pmids) > 0
RETURN m.URI AS miRNA, g.URI AS Gene, d.URI AS Disease, common_pmids AS PubMedID
\end{lstlisting}

All these queries allow the extraction of tabular data that can be used for conducting different kinds of analyses. They can be used to train supervised ML techniques or to group data according to different points of view. 

\subsection{Content-aware KG pruning and clustering}

Traditional approaches to KG pruning (i.e. the process of conflating or removing nodes and edges that are irrelevant to a specific analysis~\cite{faralli}) and clustering (i.e. grouping together highly similar nodes) typically rely on topological criteria~\cite{faralli2,tarjan} or incorporate ``domain-aware'' knowledge~\cite{kozareva,swartout}. For example, the GRAPE~\cite{grape} library supports the identification of ``topological oddities'' of a graph, that is, nodes that are topologically indistinguishable because they share the same type and neighbors. Moreover, it allows clustering together nodes and identifying communities relying on connected components, clustering coefficients, and neighborhood similarity, as well as low-dimensional projections via dimensionality reduction techniques like t-SNE.

However, in many use cases, it is desirable to complement structural information with the semantic content encoded in the node and edge attributes.
For example, ``biologically similar'' RNA molecules can be clustered by considering sequence alignment techniques~\cite{clustal,NEEDLEMAN1970443} or according to the tissue in which they are expressed or active, allowing the discovery of tissue-specific and functionally-related entities.

In all these situations, the rich set of properties that are made available in RNA-KG v2.0 can be exploited to define approaches that are both topological and content-aware. 

\begin{figure}[t]
\footnotesize 
\caption{Alignment scores distribution within sncRNA isomorphic groups.}
\centering
\begin{minipage}{0.5\linewidth} \centering 
\footnotesize
\hspace{-0.75cm}\includegraphics[width=0.61\linewidth]{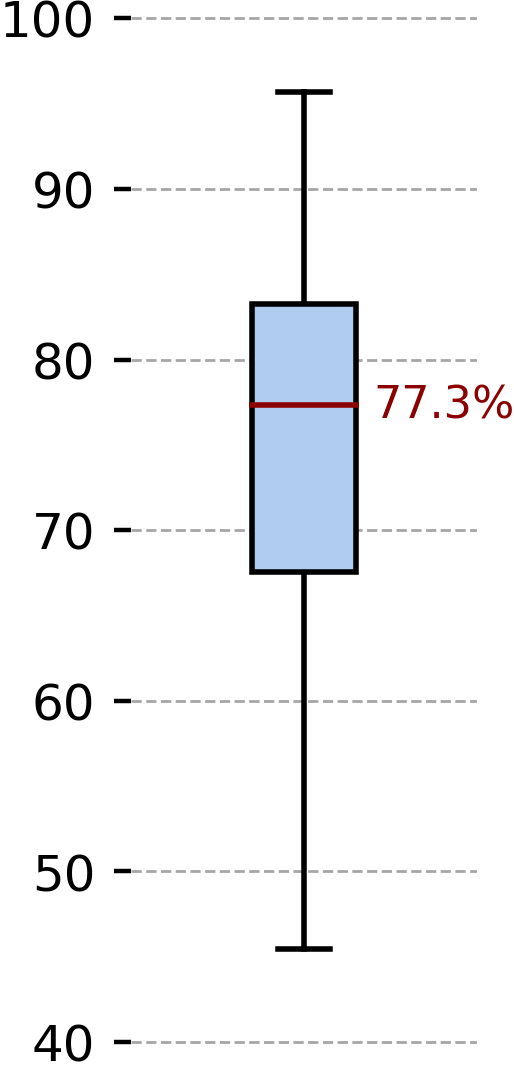}
\end{minipage}%
\begin{minipage}{0.5\linewidth} \centering 
\footnotesize
\hspace{-0.75cm}\begin{tabular}{|l|r|r|}
\hline
\begin{tabular}[x]{@{}c@{}}{\bf Alignment}\\{\bf score (\%)}\end{tabular} & \bf{Groups} & \bf{Seq.}  \\
\hline
$>$90 & 19 & 72 \\
90–80 & 47 & 386 \\
80–70 & 42 & 834 \\
70–60 & 36 & 1,182 \\
60–50 & 11 & 153 \\
$<$50 & 2 & 5 \\
        \hline
    \end{tabular}
\end{minipage}
    \label{fig:alignment}
\end{figure}

\begin{example}
Suppose we are interested in studying physical interactions involving sncRNAs. The subgraph of interest consists of approximately $9.8k$ nodes, $244.8k$ edges, and $4$ edge types: {\em interacts with}, {\em genetically interacts with}, {\em molecularly interacts with}, and {\em directly regulates activity of}. In this subgraph, we identified 177 isomorphic node groups involving $2.8k$ nodes (mainly miRNA and tsRNA sequences) and $139.3k$ edges (56.9\% of the total). 
This high density can be explained by the biological tendency of tsRNAs to act in coordinated groups~\cite{tsrna}.
A topological pruning approach would collapse all nodes in each group into a single representative, potentially reducing the edges to $48k$. Although this strategy would greatly reduce the size of the graph, it would also risk discarding biologically significant differences among the sequences.

To mitigate this phenomenon, the average global sequence alignment scores can be computed within each group using the Needleman–Wunsch algorithm~\cite{NEEDLEMAN1970443}. Fig.~\ref{fig:alignment} shows the distribution of alignment scores across groups. A total of 78 groups (49.7\%) involving 596 sequences exhibit a score exceeding the median (highlighted in red in the boxplot). 
By only collapsing nodes within groups whose members show high biological similarity (i.e. groups with alignment scores above the median), the graph size is reduced to $9.3k$ nodes and $196.3M$ edges (a 19.8\% reduction) while preserving both topological and content-aware information.
\end{example}

\subsection{Topological and context-aware link predictions}

Link prediction is a core task in KG analysis, aiming to infer new plausible edges between entities based on the existing graph structure and content. Several approaches can be applied in a KG that only exploit the network topology (homogeneous methods), the topology and the nodes' and edges' types (heterogeneous methods), or that also combine the node and edge properties (multimodal methods). 

We have recently demonstrated~\cite{torgano} that when interested in identifying the existence of an edge between two KG nodes, simple homogeneous graph representation learning (GRL) methods (e.g. LINE~\cite{line} and node2vec~\cite{node2vec}) combined with random forests achieve a balanced accuracy above 80\% for homogeneous link prediction tasks involving  miRNA-miRNA, miRNA-gene, and lncRNA-disease relationships. Thus, these approaches can be used when we are interested in the existence of a link or when the KG is quite homogeneous and a few types of edges can occur between the same pair of node types. 

In case we are interested in identifying the existence of an edge between two KG nodes and the type of relationship that exists between the nodes, heterogeneous GRL methods (e.g. TransE~\cite{transe}, ComplEx~\cite{complex}, and Metapath2Vec~\cite{meta2vec}) can be employed. These methods embed both node and edge types, thus predicting the existence of a link and the specific interaction type. 
These methods are relevant when we need to discriminate among multiple semantic relations that may exist between the same types of entities. For example, within interactions between RNAs and diseases, we might be interested in determining whether an RNA molecule is {\em over-expressed in}, {\em under-expressed in}, {\em contributes to}, or {\em causes} a given disease.

Multimodal link prediction methods~\cite{bioblp} further extend homogeneous and heterogeneous approaches by incorporating complementary information contained in nodes' and edges' attributes in the learning process. 
For instance, fine-tuned transformers like BioBERT~\cite{biobert} or DNABERT~\cite{dnabert} can be used to embed textual biomedical descriptions and sequences. These embeddings can be concatenated with those generated by GRL methods to enhance the topologically based link prediction method with biological and contextual information.

\begin{table}[t]
\footnotesize
\caption{Link prediction results on {\it miRNAdisease}.}
\centering
\begin{tabular}{ll|c|c||c|c}
\toprule
& & \multicolumn{4}{c}{\bf Balanced accuracy} \\
{\bf Model} & {\bf Edge} & {\bf Hom} & {\bf MHom} & {\bf Het} & {\bf MHet} \\
\midrule
 & Gene-Dis. & 89.09\% & {\bf 89.39\%} & 89.09\% & {\bf 89.42\%} \\
 & Gene-Phen. & 80.12\% & {\bf 80.97\%} & 80.13\% & {\bf 80.83\%} \\
LINE & RNA-Dis. & 75.38\% & {\bf 75.71\%} & 49.51\% & 49.34\% \\
 & RNA-Gene & 61.44\% & {\bf 66.49\%} & 43.58\% & {\bf 48.00\%} \\
 & RNA-Phen. & 74.29\% & {\bf 74.67\%} & 49.87\% & {\bf 50.21\%} \\
\midrule
 & Gene-Dis. & 90.43\% & 89.68\% & 90.00\% & 89.65\% \\
 & Gene-Phen. & 82.06\% & 81.31\% & 81.76\% & 81.08\% \\
n2v & RNA-Dis. & 76.28\% & 76.02\% & 50.18\% & 49.96\% \\
 & RNA-Gene & 72.31\% & {\bf 72.90\%} & 51.96\% & {\bf 52.85\%} \\
 & RNA-Phen. & 74.94\% & 74.73\% & 50.77\% & 50.33\% \\
\midrule
 & Gene-Dis. & 75.51\% & {\bf 80.47\%} & 75.53\% & {\bf 80.47\%} \\
 & Gene-Phen. & 64.30\% & {\bf 69.84\%} & 64.54\% & {\bf 69.84\%} \\
TransE & RNA-Dis. & 75.78\% & 74.41\% & 84.01\% & 53.46\% \\
 & RNA-Gene & 58.71\% & {\bf 63.99\%} & 64.87\% & {\bf 70.91\%} \\
 & RNA-Phen. & 75.85\% & 73.14\% & 83.14\% & 55.30\% \\
\bottomrule
\end{tabular}
\label{tab:lp}
\end{table}

\begin{example}
To evaluate the behavior of the different link prediction approaches on RNA-KG, we generated a {\it view} named {\it miRNAdisease}. This {\it view} has $105k$ nodes and $1.6M$ edges and aims to represent the biological and functional context of miRNAs by including their interactions with diseases, phenotypes, epigenetic changes, and genes. 
miRNAs and genes are enhanced with {\tt sequences} while diseases and phenotypes are enhanced with ontological {\tt descriptions}. Details on node distribution and edge cardinalities are reported in Supplementary Fig. S1 
(only edges with more than $1k$ occurrences are shown for the sake of readability).

Table~\ref{tab:lp} reports the balanced accuracy on the {\it miRNAdisease} {\it view}. We distinguish between homogeneous link prediction (denoted {\tt Hom}, predicting the existence of an edge, regardless of its type), heterogeneous link prediction ({\tt Het}, predicting both the existence and type of an edge), and their corresponding multimodal variants ({\tt MHom} and {\tt MHet}). DNABERT was used to embed sequences and BioBERT for descriptions. We concatenated these embeddings with those produced by the GRL methods LINE, node2vec ({\tt n2v}), and TransE.

Results confirm that node2vec outperforms LINE and TransE in homogeneous prediction; TransE outperforms LINE and node2vec in heterogeneous link prediction tasks when more than one edge type is present at schema-level (i.e. miRNA-gene and miRNA-disease). 
TransE achieves better results by disambiguating the four semantically distinct edge types between miRNAs and diseases. 
Finally, the multi-modal approach improves the performances in 9 out of 15 cases for the homogeneous link prediction task and in 8 cases for the heterogeneous task (highlighted in bold),
with gains up to +5\% in balanced accuracy observed for RNA-gene predictions in both LINE and TransE.
All experiments followed the unbiased pipeline described in~\cite{torgano} (Supplementary material 
reports the used parameters).
\end{example}

\subsection{Time-stratified link prediction}

A key property introduced in the enhanced version of RNA-KG is the PubMed identifier of scientific articles that support the existence of relations among bio-entities. From these identifiers, we can extract the publication year that can be exploited for observing the content of the KG at different instants of time. 

\begin{figure}[t]
\centering
    \includegraphics[width=.825\linewidth]{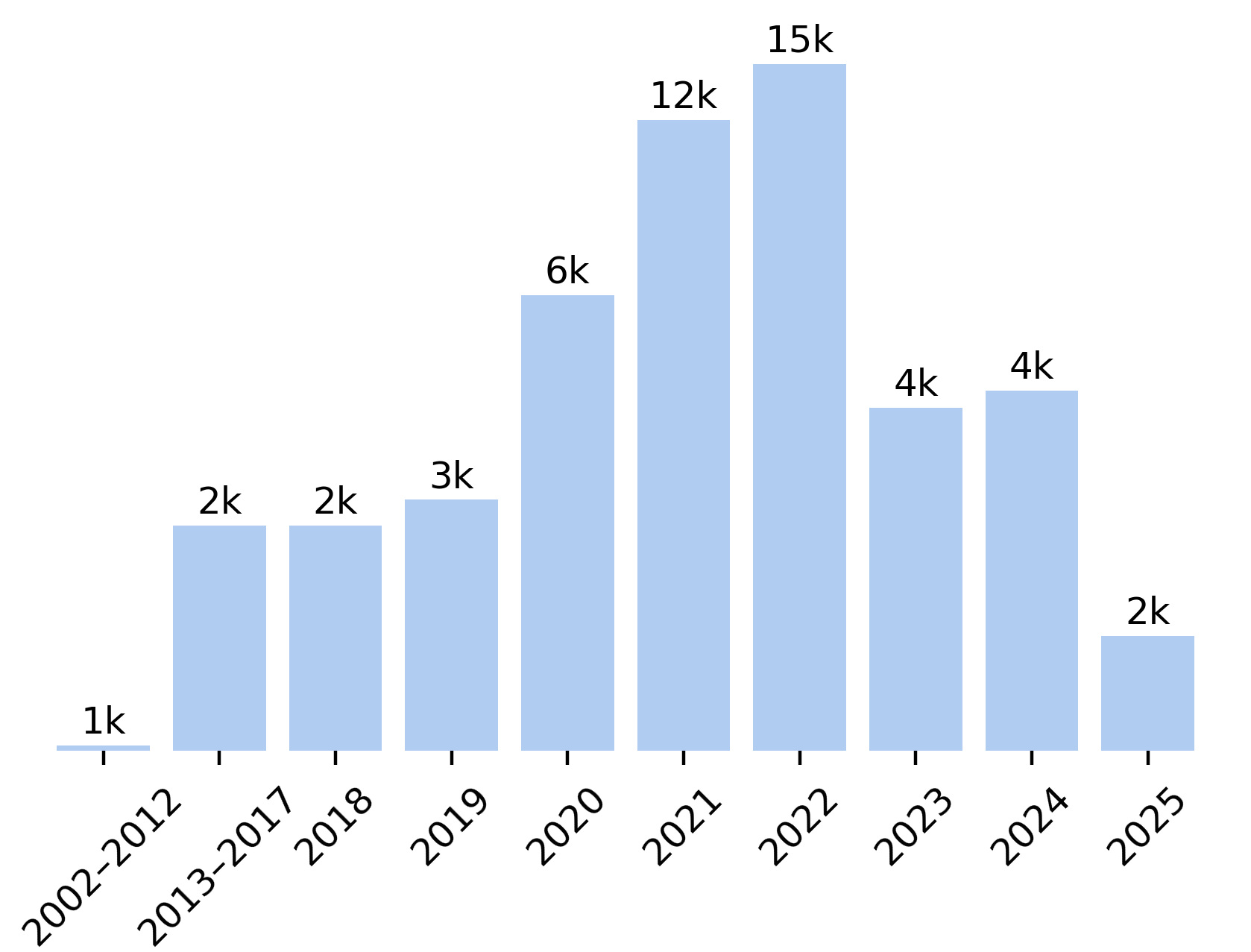}
    \caption{Distribution of articles'
    years for relations in {\it miRNAdisease}.}
    \label{fig:pubmed_year}
\end{figure}


\begin{example}
To show the relevance of the publication date to check new scientific advancements on regulatory RNA molecules for developing new therapies and vaccines, we extracted from RNA-KG interactions involving miRNA, gene, diseases, phenotypes, and epigenetic modifications. They are supported by $52k$ PubMed articles that we grouped together relying on the year of publication. 
Using this information, we can construct the distribution of articles per year as reported in Fig.~\ref{fig:pubmed_year}. Relying on this graphic, we can observe that   
more than 84\% of the relations were discovered after 2019, the year marking the emergence of COVID-19 and a subsequent surge in RNA-related studies.
\end{example}

Using the $14.3M$ edges in RNA-KG v2.0 annotated with PubMed identifiers of supporting articles, time-stratified link prediction can be realized. Models can be trained with the knowledge available up to a given time point and then evaluate their ability to predict links discovered later.
This approach can be used to validate the KG quality, as it is expected to support the discovery of future knowledge based on past evidence.

\begin{figure}[t]
    \centering
    \includegraphics[width=.975\linewidth]{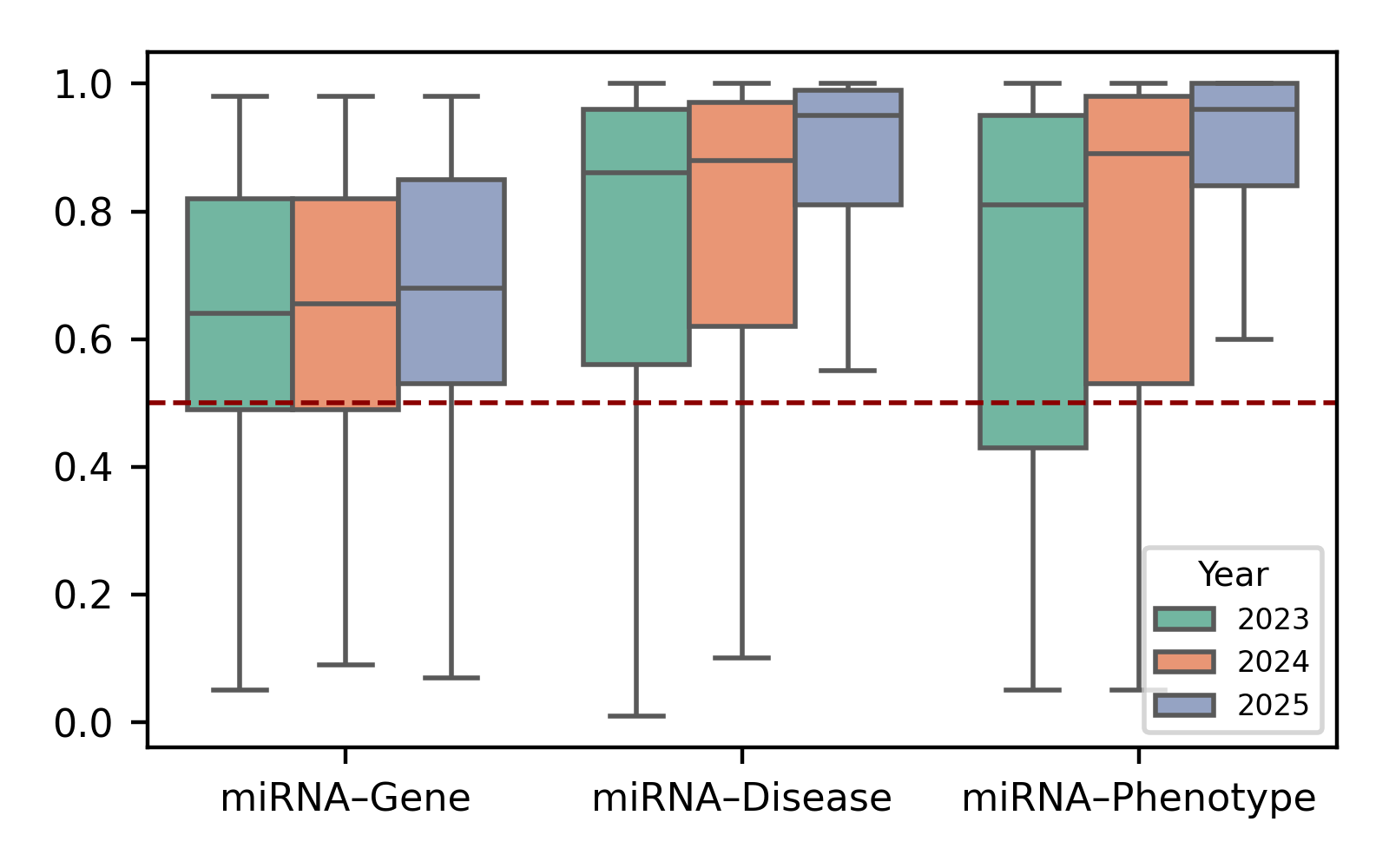}
    \caption{node2vec probability scores for relations discovered after 2022.}
    \label{fig:time_stratified_lp}
\end{figure}

\begin{example}
To validate the {\it miRNAdisease} view using a time-stratified scenario, we trained a ML model on its relations supported by articles published before 2022. Interactions discovered after 2022 are used as a test set. These links represent knowledge that was not available at training time and to which we aim to assign high probability scores.
The model has been conceived through node2vec embeddings coupled with a random forest classifier, following our usual link prediction pipeline setting~\cite{torgano}. 

Figure~\ref{fig:time_stratified_lp} reports the distributions of probability scores for interactions involving miRNA molecules in the test set. Results show that 82\% of the predicted links achieve plausibility scores well above 0.5 (highlighted in red), with median scores exceeding 0.6. 
The boxes corresponding to years 2023 and 2024 appear wider, possibly due to a larger number of novel interactions and/or increased variability in their nature.
This suggests that the KG can support meaningful predictive inference, as the model can anticipate future associations relying on prior evidence.
\end{example}



\section{Conclusion}
In this paper, we presented the new characteristics of the enhanced version of RNA-KG. It now contains properties associated with bio-entities and their interactions that can be exploited for various kinds of analysis and research activities. 
We have described various use cases that leverage the new information contained in this version of the system. The KG is publicly accessible and can be easily queried through our web portal and API.

RNA-KG can be extended in different directions. Although the KG is currently focused on human bio-entities, we would also like to include entities and relationships for other species and further information extracted by other sources. Moreover, we would like to adopt a more advanced representation based on the use of hypergraphs to better model n-ary relationships (like ceRNA) and provide easy-to-use approach for extracting {\it views} that can be exploited for ML analysis, especially link prediction techniques. We are also currently working on an integration with RNAcentral to generate a new portal, where, in addition to information about the structure of RNA molecules, the interactions in which they are involved can be made available to researchers.

The presence of an ontology is of paramount importance for the description of the current characteristics of coding and non-coding RNA molecules. However, the current tentative proposals do not take into account the heterogeneity of identification schemes across RNA sources, do not provide mapping rules between identification systems, and do not include properties that characterize RNA molecules (e.g. sequences, structures, and descriptions). Moreover, they fail to capture recent advancements in RNA research. Specifically, the Non-Coding RNA Ontology (NCRO~\cite{ncro}) was published over eight years ago and focuses on the classification of non-coding RNA molecules with an emphasis on representation of the $\sim2k$ human stem-loop miRNAs. 
However, cross-references are not up-to-date and relevant properties such as the sequence are not provided. NCRO also lacks representation of other ncRNA molecules (e.g. lncRNA, circRNA) now recognized as pivotal in therapeutic and diagnostic contexts.
The RNA Ontology (RNAO~\cite{rnao}) was published more than fifteen years ago and focuses on 3D RNA structures, including classes about biochemical and molecular types but does not include RNA sequences, limiting its utility in sequence-specific applications.

For these reasons, we would like to develop a new ontology that leverages the knowledge contained in RNA-KG v2.0 and complements existing RNA sequence repositories such as RNAcentral and Ensembl by providing an ontological representation of RNA transcripts. Look-up tables strategies~\cite{rnakg} can be used to define cross-references within the ontology.

\section{Competing interests}
No competing interest is declared.

\section{Author contributions statement}

E.C. and M.M. conceived the experiment(s),  E.C. conducted the experiment(s), E.C. and P.P. developed the RNA-KG web portal, E.C. and M.M. analysed the results. E.C., P.P., and M.M. wrote and reviewed the manuscript.

\section{Acknowledgments}
This research was in part supported by the ``National Center for Gene Therapy and Drugs based on RNA Technology'', PNRR-NextGeneration EU program [G43C22001320007], and in part by the MUSA - Multilayered Urban
Sustainability Action - Project, funded by the PNRR-Next\-Genera\-tion EU program ([G43C22001370007], Code  ECS00000037).

\bibliographystyle{plain}
\bibliography{biblio}

\clearpage

\setcounter{table}{0}
\renewcommand{\thetable}{S\arabic{table}}

\setcounter{figure}{0}
\renewcommand{\thefigure}{S\arabic{figure}}

\renewcommand{\thelstlisting}{S\arabic{lstlisting}}

\end{document}


\maketitle

\setcounter{table}{0}
\renewcommand{\thetable}{S\arabic{table}}
\renewcommand{\tablename}{Supplementary Table}

\setcounter{figure}{0}
\renewcommand{\thefigure}{S\arabic{figure}}
\renewcommand{\figurename}{Supplementary Fig.}

\renewcommand{\thelstlisting}{S\arabic{lstlisting}}
\renewcommand{\lstlistingname}{Supplementary Listing}

\vspace{-2cm}
\section*{
}

\section{
Details on sources providing additional RNA properties}\label{app}

Addgene provides gRNAs with their {\tt Cas9 species}; Apta-Index provides aptamers; MIT/ICBP siRNA includes siRNAs and shRNAs; circBase offers circRNA data with {\tt species} and {\tt genomic coordinates}; eSkip-Finder contains ASOs with {\tt species} annotations; tsRFun provides human tsRNAs; MINTBASE and tRFdb focus on tRFs, including {\tt species} and {\tt genomic coordinates}; TDBD and RSwitch deal with bacterial riboswitches, with TDBD additionally offering {\tt secondary structures}; ViroidDB provides viral RNA data from different pathogens with {\tt secondary structure} annotations; DrugBank contains ncRNA drugs (we have identified siRNAs, ASOs, and aptamers) and mRNA vaccine information, annotated with {\tt CAS numbers}, {\tt InChIKeys}, and {\tt synonyms}. Apta-Index, MIT/ICBP siRNA, and DrugBank lack species annotations due to the molecules' synthetic nature.
Moreover, although GtRNAdb's tRNA sequences are mapped to RNAcentral, they have required independent processing to extract secondary {\tt structures} (i.e. the cloverleaf structures). 

\section{
Relations and Entities integrated in RNA-KG from the new considered sources}\label{app2}

Supplementary Tables~\ref{tab:newDB1}--\ref{tab:newDB3} detail the relationships identified in the new sources that were included in RNA-KG. We also report the entities involved in these relationships. 

\begin{table*}[ht]
    \centering
    \begin{tabular}{|c|c c|c c|}
         \hline {\bf Source} & {\bf Relation} & & {\bf Entity} & \\ \hline
        DisGeNET & \begin{tabular}[x]{@{}c@{}}gene-causes or contributes to condition-disease \\ gene-causes or contributes to condition-phenotype\end{tabular} & \begin{tabular}[x]{@{}c@{}}$36k$ \\ $11k$\end{tabular} & \begin{tabular}[x]{@{}c@{}}gene\\disease\\phenotype\end{tabular} & \begin{tabular}[x]{@{}c@{}}$3k$\\$4k$\\$1k$\end{tabular}\\ \hline
        GeneMANIA & gene-genetically interacts with-gene & $2k$ & gene & $<1k$
        \\ \hline
        CTD & \begin{tabular}[x]{@{}c@{}}chemical-interacts with-GO\\GO-interacts with-chemical\\chemical-is substance that treats-disease\\disease-is treated by substance\\chemical-interacts with-protein\\protein-interacts with-chemical\\chemical-molecularly interacts with-protein\\protein-molecularly interacts with-chemical\\chemical-is substance that treats-phenotype\\phenotype-is treated by substance-chemical\\gene-participates in-pathway\\pathway-has participant-gene\\gene-causes or contributes to condition-disease\\chemical-interacts with-gene\\gene-interacts with-chemical\\gene-causes or contributes to condition-phenotype\end{tabular} & \begin{tabular}[x]{@{}c@{}}$358k$\\$358k$\\$160k$\\$160k$\\$119k$\\$119k$\\$119k$\\$119k$\\$101k$\\$101k$\\$39k$\\$39k$\\$36k$\\$12k$\\$12k$\\$11k$\end{tabular} & \begin{tabular}[x]{@{}c@{}}protein\\gene\\disease\\chemical\\GO\\pathway\\phenotype\end{tabular} & \begin{tabular}[x]{@{}c@{}}$11k$\\$7k$\\$6k$\\$6k$\\$2k$\\$2k$\\$2k$\end{tabular} \\ \hline
        ClinVar & \begin{tabular}[x]{@{}c@{}}variant-causes or contributes to condition-disease\\variant-causes or contributes to condition-phenotype\\RNA-causally influenced by-variant\\variant-causally influences-RNA\\disease-has material basis in germline mutation in-RNA\\RNA-is causal germline mutation in-disease\\phenotype-has material basis in germline mutation in-RNA\\RNA-is causal germline mutation in-phenotype\end{tabular} & \begin{tabular}[x]{@{}c@{}}$455k$\\$40k$\\$1k$\\$1k$\\$<1k$\\$<1k$\\$<1k$\\$<1k$\end{tabular} & \begin{tabular}[x]{@{}c@{}} variant\\disease\\phenotype\\RNA\end{tabular} & \begin{tabular}[x]{@{}c@{}} $259k$\\$6k$\\$<1k$\\$<1k$ \end{tabular} \\ \hline
        STRING & protein-molecularly interacts with-protein & $341k$ & protein & $13k$ \\ \hline
        starBase2 & RNA-involved in-genomic feature & $<1k$ & \begin{tabular}[x]{@{}c@{}}RNA\\genomic feature\end{tabular} & \begin{tabular}[x]{@{}c@{}}$<1k$\\$<1k$\end{tabular} \\ \hline
         microT & \begin{tabular}[x]{@{}c@{}}RNA-directly negatively regulates activity of-RNA\\RNA-directly positively regulates activity of-RNA\end{tabular} & \begin{tabular}[x]{@{}c@{}}$69k$\\$<1k$\end{tabular} & RNA & $9k$ \\ \hline
        \begin{tabular}[x]{@{}c@{}}The Human\\Protein Atlas\end{tabular} & \begin{tabular}[x]{@{}c@{}}anatomy-location of-protein\\protein-located in-anatomy\\cell-location of-protein\\protein-located in-cell\\GO-location of-protein\\protein-located in-GO\\anatomy-location of-RNA\\RNA-located in-anatomy\\cell-location of-RNA\\RNA-located in-cell\end{tabular} & \begin{tabular}[x]{@{}c@{}}$16k$\\$16k$\\$11k$\\$11k$\\$10k$\\$10k$\\$2k$\\$2k$\\$<1k$\\$<1k$\end{tabular} & \begin{tabular}[x]{@{}c@{}}protein\\RNA\\cell\\GO\\anatomy\end{tabular} & \begin{tabular}[x]{@{}c@{}}$14k$\\$1k$\\$<1k$\\$<1k$\\$<1k$\end{tabular} \\ \hline
        miRanda & \begin{tabular}[x]{@{}c@{}}RNA-involved in-genomic feature\\RNA-molecularly interacts with-RNA\end{tabular} & \begin{tabular}[x]{@{}c@{}}$29k$\\$190k$\end{tabular} & \begin{tabular}[x]{@{}c@{}}RNA\\genomic feature\end{tabular} & \begin{tabular}[x]{@{}c@{}}$92k$\\$<1k$\end{tabular} \\ \hline
        \end{tabular}
            \caption{New sources introduced in RNA-KG (I).}
    \label{tab:newDB1}
\end{table*}

\begin{table*}[ht]
    \centering
    \begin{tabular}{|c|c c|c c|}
         \hline {\bf Source} & {\bf Relation} & & {\bf Entity} & \\ \hline
        Reactome & \begin{tabular}[x]{@{}c@{}}protein-participates in-pathway\\pathway-has participant-protein\\chemical-participates in-pathway\\pathway-has participant-chemical\\GO-location of-protein\\protein-located in-GO\\GO-enabled by-protein\\protein-enables-GO\\GO-realized in response to-pathway\end{tabular} & \begin{tabular}[x]{@{}c@{}}$126k$\\$126k$\\$33k$\\$33k$\\$13k$\\$13k$\\$2k$\\$2k$\\$<1k$\end{tabular} & \begin{tabular}[x]{@{}c@{}}protein\\pathway\\chemical\\GO\end{tabular} & \begin{tabular}[x]{@{}c@{}}$11k$\\$3k$\\$2k$\\$1k$\end{tabular} \\ \hline
        HGNC & \begin{tabular}[x]{@{}c@{}}GO-enabled by-protein\\GO-location of-protein\\GO-has part-protein\\GO-not enabled by-protein\\GO-colocalizes with-protein\\GO-not location of-protein\\protein-enables-GO\\protein-located in-GO\\protein-part of-GO\\protein-contributes to-GO\\protein-not enables-GO\\protein-not involved in-GO\\protein-colocalizes with-GO\\protein-is active in-GO\\protein-not located in-GO
        \end{tabular} & \begin{tabular}[x]{@{}c@{}}$<1k$\\$<1k$\\$<1k$\\$<1k$\\$<1k$\\$<1k$\\$<1k$\\$<1k$\\$<1k$\\$<1k$\\$<1k$\\$<1k$\\$<1k$\\$<1k$\\$<1k$\end{tabular} & \begin{tabular}[x]{@{}c@{}}protein\\GO\end{tabular} & \begin{tabular}[x]{@{}c@{}}$<1k$\\$<1k$\end{tabular} \\ \hline
        UniProtKB & \begin{tabular}[x]{@{}c@{}}GO-enabled by-protein\\GO-location of-protein\\protein-enables-GO\\protein-located in-GO\\GO-has part-protein\\protein-part of-GO\\chemical-molecularly interacts with-protein\\protein-molecularly interacts with-chemical\\protein-is active in-GO\\GO-colocalizes with-protein\\GO-not enabled by-protein\\GO-not location of-protein\\GO-not has part-protein\\GO-not colocalizes with-protein\\protein-colocalizes with-GO\\protein-not involved in-GO\\protein-contributes to-GO\\protein-not enables-GO\\protein-not located in-GO\\protein-acts upstream of or within-GO\\protein-acts upstream of positive effect-GO\\protein-acts upstream of or within positive effect-GO\\protein-acts upstream of negative effect-GO\\protein-not part of-GO\\protein-not colocalizes with-GO\\protein-acts upstream of or within negative effect-GO\\protein-not is active in-GO\\RNA-enables-GO\\GO-enabled by-RNA
        \end{tabular} & \begin{tabular}[x]{@{}c@{}}$41k$\\$41k$\\$33k$\\$33k$\\$20k$\\$20k$\\$18k$\\$18k$\\$1k$\\$<1k$\\$<1k$\\$<1k$\\$<1k$\\$<1k$\\$<1k$\\$<1k$\\$<1k$\\$<1k$\\$<1k$\\$<1k$\\$<1k$\\$<1k$\\$<1k$\\$<1k$\\$<1k$\\$<1k$\\$<1k$\\$<1k$\\$<1k$\end{tabular} & \begin{tabular}[x]{@{}c@{}}protein\\GO\\chemical\\RNA\end{tabular} & \begin{tabular}[x]{@{}c@{}}$18k$\\$5k$\\$2k$\\$<1k$\end{tabular} \\ \hline
            \end{tabular}
    \caption{New sources introduced in RNA-KG (II).}
    \label{tab:newDB2}
\end{table*}

\begin{table*}[ht]
    \centering
    \begin{tabular}{|c|c c|c c|}
         \hline {\bf Source} & {\bf Relation} & & {\bf Entity} & \\ \hline
        PhenomiR & \begin{tabular}[x]{@{}c@{}}RNA-over-expressed in-disease\\RNA-under-expressed in-disease\\RNA-over-expressed in-phenotype\\RNA-under-expressed in-phenotype\\RNA-causes or contributes to condition-disease\\RNA-causes or contributes to condition-phenotype\end{tabular} & \begin{tabular}[x]{@{}c@{}}$3k$\\$3k$\\$1k$\\$1k$\\$<1k$\\$<1k$\end{tabular} & \begin{tabular}[x]{@{}c@{}}RNA\\disease\\phenotype\end{tabular} & \begin{tabular}[x]{@{}c@{}}$<1k$\\$<1k$\\$<1k$\end{tabular} \\ \hline
        COSMIC & \begin{tabular}[x]{@{}c@{}}RNA-causally influenced by-variant\\variant-causally influences-RNA\end{tabular} & \begin{tabular}[x]{@{}c@{}}$738k$\\$738k$\end{tabular} & \begin{tabular}[x]{@{}c@{}}variant\\RNA\end{tabular} & \begin{tabular}[x]{@{}c@{}}$566k$\\$13k$\end{tabular} \\ \hline
        RNAcentral & \begin{tabular}[x]{@{}c@{}}RNA-in similarity relationship with-RNA\\RNA-located in-GO\\GO-location of-RNA\\RNA-participates in-GO\\GO-has participant-RNA\\RNA-part of-GO\\GO-has part-RNA\\RNA-has function-GO\\GO-function of-RNA\\RNA-enables-GO\\GO-enabled by-RNA\end{tabular} & \begin{tabular}[x]{@{}c@{}}$4M$\\$19k$\\$19k$\\$17k$\\$17k$\\$10k$\\$10k$\\$10k$\\$10k$\\$10k$\\$10k$        \end{tabular} & \begin{tabular}[x]{@{}c@{}}RNA\\GO\end{tabular} & \begin{tabular}[x]{@{}c@{}}$25k$\\$<1k$\end{tabular} \\ \hline
        circBase & \begin{tabular}[x]{@{}c@{}}RNA-located in-cell\\cell-location of-RNA\\RNA-transcribed from-gene\\gene-transcribed to-RNA\end{tabular} & \begin{tabular}[x]{@{}c@{}}$68k$\\$68k$\\$32k$\\$32k$\end{tabular} & \begin{tabular}[x]{@{}c@{}}RNA\\gene\\cell\end{tabular} & \begin{tabular}[x]{@{}c@{}}$61k$\\$4k$\\$<1k$\end{tabular} \\ \hline
        Ensembl & \begin{tabular}[x]{@{}c@{}}gene-transcribed to-RNA\\RNA-transcribed from-gene\\RNA-ribosomally translates to-protein\\protein-robosomal translation of-RNA\\protein-gene product of-gene\\gene-has gene product-protein\\GO-location of-protein\\protein-located in-GO\\GO-enabled by-protein\\protein-enables-GO\\GO-has part-protein\\protein-part of-GO\end{tabular} & \begin{tabular}[x]{@{}c@{}}$122k$\\$122k$\\$11k$\\$11k$\\$8k$\\$8k$\\$5k$\\$5k$\\$3k$\\$3k$\\$<1k$\\$<1k$\end{tabular} & \begin{tabular}[x]{@{}c@{}}RNA\\gene\\protein\\GO\end{tabular} & \begin{tabular}[x]{@{}c@{}}$124k$\\$14k$\\$13k$\\$1k$\end{tabular} \\ \hline
        RNAhybrid & RNA-molecularly interacts with-RNA & $189k$ & RNA & $71k$ \\ \hline
        GTEx & \begin{tabular}[x]{@{}c@{}}RNA-located in-anatomy\\anatomy-location of-RNA\end{tabular} & \begin{tabular}[x]{@{}c@{}}$322k$\\$322k$\end{tabular} & \begin{tabular}[x]{@{}c@{}}RNA\\anatomy\end{tabular} & \begin{tabular}[x]{@{}c@{}}$59k$\\$<1k$\end{tabular} \\ \hline
        POSTAR2 & RNA-involved in-genomic feature & $29k$ & \begin{tabular}[x]{@{}c@{}}RNA\\genomic feature\end{tabular} & \begin{tabular}[x]{@{}c@{}}$21k$\\$<1k$\end{tabular} \\ \hline
    \end{tabular}
    \caption{New sources introduced in RNA-KG (III).}
    \label{tab:newDB3}
\end{table*}

\section{
Link prediction setup}\label{appc}

\begin{figure}[ht]
\footnotesize 
\caption{{\it miRNAdisease} degree distribution and node and edge cardinalities.}
\centering
\includegraphics[width=0.5\linewidth]{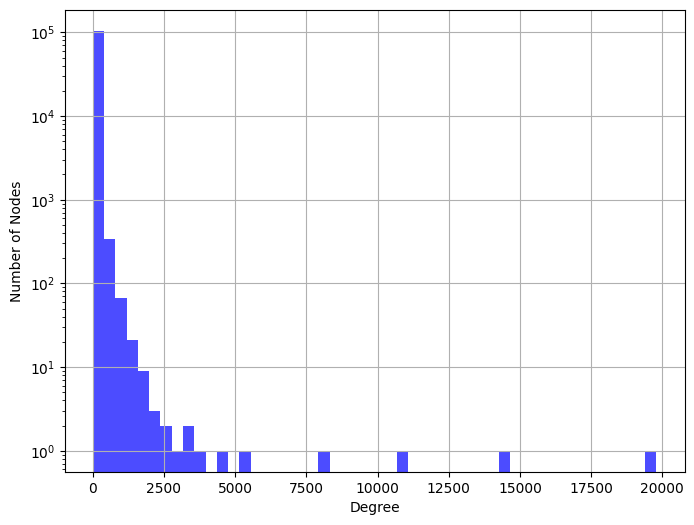}
\vspace{0.25cm}\begin{minipage}{0.7\linewidth} \centering 
\begin{tabular}{|l|r|}
\hline
\bf{Node} & \bf{Count}  \\
\hline
Gene & 48,395 \\ 
Disease & 26,015 \\ 
Phenotype & 19,025 \\ 
miRNA & 9,114 \\ 
Genomic feature & 2,393 \\ 
        \hline
    \end{tabular}
\end{minipage}%

\footnotesize
\begin{minipage}{\linewidth} \centering
\begin{tabular}{|l|l|r|l|}
\hline
\bf{Source} & \bf{Target} & \bf{Count} & \bf{Edge type(s)}  \\
\hline
Disease & Phenotype & 514,916 & \begin{tabular}[c]{@{}l@{}}has phenotype\end{tabular}\\ \hline 

Phenotype & Disease & 514,916 & \begin{tabular}[c]{@{}l@{}}phenotype of\end{tabular}\\ \hline 

Gene & Disease & 104,720 & \begin{tabular}[c]{@{}r@{}}causes or contributes to condition	\end{tabular}\\ \hline 

miRNA & miRNA & \begin{tabular}[c]{@{}r@{}} 100,926 \\ 2,810 \\ 2,810\end{tabular} & \begin{tabular}[c]{@{}l@{}}in similarity relationship with \\ develops into \\ develops from\end{tabular}\\ \hline 

miRNA & Gene & \begin{tabular}[c]{@{}r@{}}67,954 \\ 1,843\end{tabular} & \begin{tabular}[c]{@{}l@{}}regulates activity of \\ transcribed from\end{tabular}\\ \hline 

Gene & Genomic feature & 48,399 & \begin{tabular}[c]{@{}l@{}}subClassOf\end{tabular}\\ \hline 

Disease & Disease & 44,496 & \begin{tabular}[c]{@{}l@{}}subClassOf\end{tabular}\\ \hline 

miRNA & Disease & \begin{tabular}[c]{@{}r@{}}36,262 \\ 7,287 \\ 5,889 \\ 1,356\end{tabular} & \begin{tabular}[c]{@{}l@{}}causes or contributes to condition \\ under expressed in \\ over expressed in \\ is causal germline mutation in\end{tabular}\\ \hline

Gene & Phenotype & 33,228 & \begin{tabular}[c]{@{}l@{}}causes or contributes to condition	\end{tabular}\\ \hline  

Phenotype & Phenotype & 24,969 & \begin{tabular}[c]{@{}l@{}}subClassOf\end{tabular}\\ \hline

miRNA & Genomic feature & 14,627 & \begin{tabular}[c]{@{}l@{}}subClassOf\end{tabular}\\ \hline  

miRNA & Phenotype & \begin{tabular}[c]{@{}r@{}}9,181 \\ 3,266 \\ 2,511\end{tabular} & \begin{tabular}[c]{@{}l@{}}causes or contributes to condition \\ under expressed in \\ over-expressed in\end{tabular}\\ \hline  

Gene & Gene & 3,615 & \begin{tabular}[c]{@{}l@{}}genetically interacts with\end{tabular}\\ \hline  

Genomic feature & Genomic feature & 2,520 & \begin{tabular}[c]{@{}l@{}}subClassOf\end{tabular}\\ \hline  

Gene & miRNA & 1,843 & \begin{tabular}[c]{@{}l@{}}transcribed to\end{tabular}\\ \hline  

Disease & miRNA & 1,356 & \begin{tabular}[c]{@{}l@{}}has material basis in\\germline mutation in\end{tabular}\\ \hline  

    \end{tabular}
    \end{minipage}
    \label{fig:statsView4}
\end{figure}

Supplementary Fig.~\ref{fig:statsView4} provides the node distribution and node and edge cardinalities of {\it miRNAdisease}. Only edges with more than $1k$ occurrences are displayed for the sake of readability.

The following table defines the link prediction experimental setup used in the unbiased pipeline
.\\

\begin{tabular}{l|l}
\toprule
{\bf Parameter} & {\bf Value} \\
\midrule
edge\_embedding\_methods & Concatenate \\ 
use\_scale\_free\_distribution & True \\
training\_unbalanced\_rate & 1 \\
max\_depth & 100 \\
n\_estimators & 100 \\
n\_jobs & -1 \\
evaluation\_schema & Connected Monte Carlo \\
use\_scale\_free\_distribution & True \\
validation\_unbalance\_rate & 1 \\
train\_size & 0.7 \\
number\_of\_holdouts & 5 \\
\bottomrule
\end{tabular}

\section{
RNA-KG v2.0 API endpoints}\label{appd}

Supplementary Table~\ref{app:api_table} provides a summary of the RNA-KG v2.0 main API endpoints. For each endpoint, we report a short description that includes the input parameters, the HTTP method ({\em POST}/{\em GET}), and examples of input and output. Endpoints are also defined for selectively retrieving incoming or outgoing edges for a node. The complete API documentation with examples is available at \url{https://RNA-KG.biodata.di.unimi.it/api/v1/docs}.

\begin{table}[ht]
\centering
\begin{footnotesize}
\begin{tabular}{|l|p{1.5cm}|c|p{3.1cm}|p{7.5cm}|}
\hline
\textbf{Endpoint} & \textbf{Description} & \textbf{Method} & \textbf{Input example} & \textbf{Output example} \\
\hline



\texttt{/query/} & Execute a custom Cypher query on RNA-KG v2.0 & POST & 
\begin{lstlisting}[style=json]
{
  "query": "MATCH (n) RETURN n.URI LIMIT 3"
}
\end{lstlisting}
& 
\begin{lstlisting}[style=json]
{
    "results": [
        { "n.URI": "https://rnacentral.org/rna/
                    URS0000019DAD_9606"},
        { "n.URI": "https://rnacentral.org/rna/
                    URS0000019F30_9606" },
        { "n.URI": "https://rnacentral.org/rna/
                    URS0000019F58_9606" }
    ]
}
\end{lstlisting}
\\
\hline

\texttt{/node/id} & Retrieve all properties for a node given its identifier in RNA-KG v2.0 & GET & {\tt node\_id= URS00005F5B9E\_9606\& node\_id\_scheme= RNAcentral} & 
\begin{lstlisting}[style=json]
{
    "node_uri": "https://rnacentral.org/rna/
                 URS00005F5B9E_9606",
    "node_id": "RNAcentral:URS00005F5B9E_9606", 
    "node_labels": [ "RNA", "sncRNA",
                    "Small_regulatory_ncRNA",
                    "ncRNA", "miRNA" ],
    "node_properties": { 
        "Description": "homo sapiens (human)
                        hsa-mir-106a-3p", 
        "Label": "hsa-miR-106a-3p", 
        "Genomic_coordinates":
            ["chrX:134170208-134170229-"], 
        "Sequence": "CUGCAAUGUAAGCACUUCUUAC", 
        "Species": "Homo sapiens"
    }
}
\end{lstlisting}
\\ \hline

\texttt{/relationships/id} & Retrieve all relationships (incoming and outgoing) for a node given its identifier in RNA-KG v2.0 & GET & {\tt node\_id= URS00000537B8\_9606\& node\_id\_scheme= RNAcentral} & \begin{lstlisting}[style=json]
{
    "relationships": [
    {
        "relationship_type": "interacts_with",
        "relationship_properties": {
            "Source": [ "LncRNAWiki" ],
            "Context": [ "hela cell" ],
            "Method": [ "cross-linking and
                         immunoprecipitation" ],
            "PubMedID": [ "25531890" ]
        },
        "node_uri": "https://rnacentral.org/rna/
                     URS00005F5B9E_9606",
        "node_id": "RNAcentral:URS00005F5B9E_9606",
        "node_labels": [ "RNA", "sncRNA", ..., 
                         "ncRNA", "miRNA" ],
        "node_properties": { 
            "Description": "homo sapiens (human)
                            hsa-mir-106a-3p",
            "Label": "hsa-miR-106a-3p",
            "Genomic_coordinates":
                [ "chrX:134170208-134170229-" ],
            "Sequence": "CUGCAAUGUAAGCACUUCUUAC",
            "Species": "Homo sapiens"
        }
     }, ...
}
\end{lstlisting}
\\ \hline



\texttt{/rel\_metadata} & Get metadata (properties, count) for a given relationship type & GET & {\tt rel\_type= develops\_from} & 
\begin{lstlisting}[style=json]
{
  "relationship_type": "develops_from",
  "properties": [ "Source" ],
  "total_count": 36945
}
  \end{lstlisting}
 \\ \hline

\end{tabular}
\end{footnotesize}
\caption{Summary of the RNA-KG v2.0 API main endpoints.}
\label{app:api_table}
\end{table}